\documentstyle[rotate,psfig]{mn}
\newcommand{\bugref}{\bibitem[\protect\citename{dummy }1893]{dum}}
\newcommand{\etal}{et~al.\ }

\begin{document}

\title[Polarization of BL Lac Objects]
{Analysis of $\lambda = 6$~cm VLBI polarization observations of a complete 
sample of northern BL~Lacertae Objects}
\author[D.~C.~Gabuzda, A.~B.~Pushkarev and T.~V.~Cawthorne.]
{D.~C.~Gabuzda$^{1,2}$, A.~B.~Pushkarev$^{2}$ and T.~V.~Cawthorne.$^{3}$ \\
$^{1}$Joint Institute for VLBI in Europe, Postbus 2, 7990 AA Dwingeloo,
The Netherlands\\
$^{2}$Astro Space Centre, P.N. Lebedev Physical Institute, Leninsky
Prospekt 53, 117924 Moscow, Russia.\\
$^{3}$Department of Physics and Astronomy, University 
of Central Lancashire, Preston, Lancashire, PR1 2HE.\\
}

\maketitle
\begin{abstract}
The results of VLBI total intensity ($I$) and linear polarization ($P$) 
observations at $\lambda = 6$~cm are presented for ten radio bright BL Lacertae 
objects. These images complete first-epoch 
polarization observations for the 1-Jy sample of northern BL Lacertae 
objects defined by K\"uhr and Schmidt. Estimates of superluminal speeds are
presented for several sources, bringing the total number of sources in the 
sample for which such estimates are available to 16. Second epoch observations
currently being reduced should yield speed estimates for VLBI features in
essentially all the sources in the sample. The jet magnetic fields of these
BL Lacertae objects are usually transverse to the local jet direction, but a 
sizeable minority (about 30\%) have VLBI jet components with longitudinal 
magnetic fields. This may suggest that the conditions in the VLBI jets
of BL Lacertae objects are favorable for the formation of relativistic
shocks; alternatively, it may be that the toroidal component of the intrinsic
jet magnetic field is characteristically dominant in these sources. 
\end{abstract}
\begin{keywords}
BL Lacertae objects: general --
BL Lacertae objects: individual: (0003--066, 0814+425, 0820+225, 0823+033, 
1334--127, 1732+389, 2131--021, 2150+173, 2155--152, 2254+074 --
polarization --- radio sources: galaxies
\end{keywords}

\section{Introduction}

BL~Lacertae objects are active galactic nuclei whose most characteristic
distinguishing property is their relatively low-luminosity optical line 
emission; in many cases, their optical continua are completely featureless. 
Like many high-polarization quasars, BL Lacertae objects have 
strong and variable polarization in wavebands ranging from
optical through radio; they usually have compact, flat-spectrum radio
structure, and point-like optical structure. For some, luminous elliptical host
galaxies are observed (Angel and Stockman 1980; Miller 1981; Kollgaard 1994),
though the optical images of many BL Lacertae objects remain unresolved, even
in high-resolution observations (e.g. Falomo 1996). The radio emission 
and much of the optical 
emission is believed to be synchrotron radiation. Historically, BL Lacertae 
objects were first detected primarily via radio surveys, and strong radio 
emission was earlier thought to be characteristic of this type of AGN. 
More recently, large numbers of BL Lacertae objects with much weaker radio 
emission have been discovered by X-ray surveys.  The relationship 
between these so-called "radio" and X-ray" BL Lacertae objects 
is not entirely clear; the most popular current hypotheses are (1) that they are
similar objects whose spectral energy distributions peak in the infrared and
X-ray, respectively, with the radio emission in radio BL Lacertae 
objects experiencing a larger intrinsic relativistic enhancement 
(Giommi \& Padovani 1994; Padovani \& Giommi 1995; Fossati \etal 1997), 
and (2) that they are intrinsically 
identical objects whose jets are oriented at different characteristic angles 
to the line of sight, with X-ray BL Lacertae objects being viewed further 
``off-axis'' (Stocke \etal 1985; Maraschi \etal 1986). 
In the broader context of unified schemes, it is usually thought that 
the ``parent population'' of BL Lacertae objects is primarily FR~I radio 
galaxies (Browne 1983; Wardle, Moore \& Angel 1984).

Previous VLBI polarization observations of radio BL Lacertae objects at 
6~cm and 3.6~cm (Gabuzda \& Cawthorne 1996, Gabuzda \etal 1999, and references 
therein) have revealed
a tendency for the electric vector $\chi$ in polarized knots in the VLBI jets
to lie nearly along the local jet direction. 
The degrees of polarization in the jet components of BL Lacertae objects have
been observed to be as high as $m\sim 60-70\%$, with typical values $m\sim
5-15\%$, indicating that these components are optically thin and that in
at least some cases the magnetic field is very highly ordered. Assuming the
jet components to be optically thin, the observed typical $\chi$
orientation implies that the associated magnetic fields are 
perpendicular to the direction of the jet. One natural interpretation of 
this transverse magnetic field structure is that the visible jet components 
are associated with relativistic shocks that compress an initially tangled 
magnetic field, enhancing the magnetic field transverse to the compression 
(Laing 1980; Hughes, Aller and Aller 1989). As images for more BL Lacertae 
objects have become available, it has become increasingly clear 
that there are also a sizeable minority of these sources in which, in 
contrast, longitudinal magnetic fields dominate in at least some parts of 
their VLBI jets. 
\begin{table}
\centering
\def\baselinestretch{1}
\caption[xx]{{\sc Antennas Used}}
\label{tab:ants}
\vspace*{0.25in}
\small
\begin{tabular}{lccccc} \hline\hline
Antenna & Diameter & T$_{sys, R}$ & T$_{sys, L}$ & $g$ & Epoch\\
        &  (m)     & (K)  & (K) & (K/Jy) \\ \hline
Medicina      & 32  & 60 & 50 & 0.160 & 1, 2\\
Effelsberg    & 100 & 60 & 50 & 0.991 & 1, 2 \\
WSRT          & $\sqrt{14}\times$25  & 120 & 120 &       & 2\\
St. Croix     & 25  & 45 & 35 & 0.102 & 2\\
Hancock       & 25  & 55 & 60 & 0.092 & 2\\
Haystack      & 37  & 75 & 65 & 0.130 & 1\\
Green Bank    & 43  & 35 & 40 & 0.227 & 1, 2\\
North Liberty & 25  & 45 & 45 & 0.120& 1, 2\\
Phased VLA    & $\sqrt{27}\times$25 & ** & ** & ** & 1\\
Kitt Peak     & 25  & 65 & 65 &  0.120 & 1\\
Owens Valle   & 25  & 110 & 110 & 0.120& 1, 2\\
Brewster      & 25  & 35 & 35 & 0.088 & 2\\
Mauna Kea     & 25  & 35 & 35 & 0.090 & 2\\
\multicolumn{6}{l}{Epochs: 1 = 1992.23; 2 = 1995.41}\\
\end{tabular}
\end{table}

At 6~cm, the degrees of polarization of the cores of BL Lacertae 
objects ($\sim 2-9\%$) are, on average, higher than those of quasars 
($\leq 2\%$). Recent 6~cm space VLBI polarization observations of the 
BL Lacertae object 1803+784 (Gabuzda 1999) indicate that this is due 
to the fact that the observed core 
polarizations include a substantial contribution from newly emerging knots, 
as suggested by Gabuzda \etal (1994b).  Previous observations have shown 
the distribution of core $\chi$ orientations at 6~cm to be bimodal, with 
$\chi$ either aligned with or transverse to the inner jet direction; 
there is some evidence that 
$\chi_{core}$ is roughly perpendicular to the jet when the cores 
are quiescent, and aligns with the jet at epochs when emission from newly 
emerging shock components dominate the observed ``core'' polarization 
(Gabuzda \etal 1994b, Gabuzda \& Cawthorne 1996).

To elucidate the nature of the characteristic VLBI total intensity ($I$) and 
polarization ($P$) structures observed for radio-bright BL Lacertae objects, 
we are engaged in an ongoing project to obtain multi-epoch, multi-frequency 
VLBI $I$ and $P$ images for all sources in the complete sample of BL Lacertae 
objects defined by K\"uhr and Schmidt (1990).  These sources have 6~cm 
fluxes of at least 1 Jy, radio spectral indices $\alpha \geq -0.5$
($S_{\nu}\sim\nu^{+\alpha}$), rest frame equivalent widths of the strongest 
emission lines less than 5~\AA, and optical counterparts on the Palomar Sky 
Survey plates with brightness greater than 20$^m$. VLBI $I$ and $P$ images 
for most of the sample sources are presented by Roberts, Gabuzda \& Wardle 
(1987); Gabuzda, Roberts \& Wardle (1989b); Gabuzda \etal (1992, 1994b); and 
Gabuzda, Pushkarev \& Cawthorne (1999).  The ten sets of $I$ and 
$P$ images presented here complete our first-epoch images for the 
34 sources in the original sample (one, 1334--127, was subsequently
reclassified as a quasar, so that the final number of sources in the
complete BL Lac sample is 33). These 
are the first $P$ images for these ten sources, and in a number of cases, 
the first 6~cm $I$ images as well. 

\section{Observations}
 
\begin{table}
\centering
\def\baselinestretch{1}
\caption[xx]{{\sc Integrated Rotation Measures}}
\vspace*{0.25in}
\label{tab:rm}
\begin{tabular}{lccc}
 \hline\hline
Source     & RM            & RM Ref & Rot. at $\lambda$6~cm \\
           & (rad/m$^{2}$) &           & (deg) \\ \hline
0003$-$066 &   +13         &     2     & \phantom{0}+3 \\
0814+425   &   +23         &     1     & \phantom{0}+5 \\
0820+225   &   +81         &     2     & +17\\
0823+033   & \phantom{+1}1 &     2     & \phantom{0}+1  \\
1334$-$127 &   $-23$       &     2     & \phantom{0}$-5$ \\
1732+389   & +67           &     2     &  +14       \\
2131$-$021 & +46           &     2     & \phantom{0}+9 \\
2150+173   & $-39$         &     2     & \phantom{0}-8 \\
2155$-$152 & +19           &     2     & \phantom{0}+4  \\
2254+074   & $-30$         &     2     & \phantom{0}$-6$  \\
\multicolumn{4}{l}{References: 1 = Rusk 1988;}\\
\multicolumn{4}{l}{2 = Gabuzda and Pushkarev, in preparation.}\\
\end{tabular}
\end{table}
The observations of all sources except 0003--066 were made in March 1992 
(1992.23), using an 8-element global VLBI array.  The observations for 
0003--066 were made in May 1995 (1995.41) using a 10-element global array.
The antennas used in both experiments are listed in Table~\ref{tab:ants}. 
The observations were made under the auspices of the US and European VLBI 
networks. The data were recorded using the MkIII system, and the data were 
subsequently correlated using the Mk~IIIA correlators at Haystack Observatory
(1992.23) and the Max-Planck-Institut-f\"ur-Radioastronomie in Bonn (1995.41). 

The polarization calibration of these data was performed as described by
Roberts, Wardle and Brown (1994). The instrumental polarizations (``D-terms'')
were determined using observations of the unpolarized sources 3C84 and
OQ208; the overall D-terms residuals are at a level $\simeq 0.5\%$. The 
VLBI polarization position angles were calibrated by comparing the total
VLBI-scale and VLA core polarizations for the especially compact and
relatively strongly polarized sources 0823+033 and 1749+096. The rotations 
required to align $\chi$ for the total VLBI-scale polarizations of these
sources with $\chi$ for their VLA core polarizations agreed to within
less than one degree, suggesting that the overall VLBI polarization position
angle calibration is good to about this level of precision. 

Images of the distribution of total
intensity $I$ were made using a self-calibration algorithm similar to that
described by Cornwell and Wilkinson (1981). Maps of the linear
polarization\footnote{$P = pe^{2i\chi} = mIe^{2i\chi}$, where $p=mI$ is the
polarized intensity, $m$ is the fractional linear polarization, and $\chi$ is
the position angle of the electric vector on the sky, measured from north
through east.} $P$ were made by
referencing the calibrated cross-hand fringes to the parallel-hand fringes
using the antenna gains determined in the hybrid mapping, Fourier transforming
the cross-hand fringes, and performing a complex CLEAN. One byproduct of this
procedure is to register the $I$ and $P$ maps to within a small fraction of a
beamwidth, so that corresponding $I$ and $P$ images may be directly
superimposed.

\section{Results}
 
\begin{table*}
\centering
\def\baselinestretch{1}
\caption[xx]{{\sc Source Models}}
\label{tab:models}
\begin{tabular}{cccccccccc} \hline \hline
   & $I$ & $p$ & {$\chi_{0}$}$^{a}$ & $m$ & $r$ & $\Delta r$ & $\theta$ & $\Delta\theta$ & FWHM \\ 
   & (mJy) & (mJy) & (deg) & (\%) & (mas)& (mas) & (deg) & (deg) & (mas)  \\ \hline
\multicolumn{10}{c}{0003$-$066}\\
 C   & 922 & 63.3  & 79  & 6.8    & -- & --& -- & -- & 0.25    \\
 K6  & 824 & $<10$ & --  & $<1.2$ & 0.60 &      & $-40$ &   & 0.23\\
 K5  & 372 & 55.8  & 21  & 13.6   & 1.46 &      & $-67$ &   & 0.46\\
 K4  & 150 & $<10$ & --  & $<7$   & 2.51 &      & $-70$ &   & 1.94\\
 K3  &  11 & $<10$ & --  & $<7$   & 4.31 &      & $-68$ &   & 0.02\\
 K2  &  15 & $<10$ & --  & $<38$  & 5.23 &      & $-66$ &   & 0.33\\ 
 K1  &  97 & $<10$ & --  & $<11$  & 6.10 &      & $-74$ &   & 1.15 \\ \hline
\multicolumn{10}{c}{0814+425}\\
 C   &  620 & $<3$ & -- & $<0.5$ & -- & -- & -- & -- & 0.48\\ 
 K5  &  200 & 22.2 & 26 & 11.1   & 0.99 & 0.01 & 128 & 1 & 0.50\\ 
 K4  &   35 &  9.2 & 84 & 26.2   & 2.29 & 0.02 & 111 & 1 & 0.92\\ 
 K3  &   11 & $<3$ & -- & $<27$  & 3.56 & 0.03 & 108 & 1 & 0.74\\ 
 K2  &   19 & $<3$ & -- & $<16$  & 4.89 & 0.07 & 101 & 2 & 1.48\\ 
 K1  &   10 & $<3$ & -- & $<30$  & 6.34 & 0.13 & 106 & 2 & 1.08\\ \hline
\multicolumn{10}{c}{0820+225}\\
 C   & 198 & $<3$  & -- & $<1.5$& --   & --   & -- & -- & 0.20\\
 K11 & 119 & 7.4   & $-26$ & 6.2   & 2.16 & 0.07 & $-173$ & 1 & 0.80\\
 K10 &  43 & 8.0   & $-69$ & 18.6  & 4.26 & 0.17 & $-167$ & 1 & 1.03\\
 K9  &  47 & $<3$  & --    & $<6.4$& 5.84 & 0.13 & $-156$ & 1 & 1.51\\
 K8  &  68 & 12.2  & 16    & 17.9  & 8.86 & 0.09 & $-133$ & 1 & 2.28\\
 K7  &  22 & 9.2   & $-5$  & 41.8  &11.33 & 0.16 & $-132$ & 1 & 1.52\\
 K6  & 113 & 7.2   & $-16$ & 6.4   &12.84 & 0.07 & $-130$ & 1 & 2.95\\
 K5  &  90 & 20.8  & $-29$ & 23.1  &17.45 & 0.08 & $-132$ & 1 & 2.87\\
 K4  &  13 & $<3$  & --    & $<23$ &19.99 & 0.10 & $-136$ & 1 & 0.24\\
 K3  & 165 & 24.1  & $-70$ & 14.6  &21.70 & 0.06 & $-135$ & 1 & 1.76\\
 K2  &  60 & 9.7   & $-80$ & 16.2  &23.38 & 0.08 & $-132$ & 1 & 1.27\\
 K1  & 180 & $<3$  & --    & $<1.7$&24.81 & 0.07 & $-138$ & 1 & 2.69\\ \hline 
\multicolumn{10}{c}{0823+033}\\
 C   & 1494& 60.6  & $-13$ & 4.0  & -- & -- & -- & -- & 0.07\\
 K3  &  415& $<15$ & -- & $<3.6$& 0.71 & 0.03 & 30 & 3 & 0.18\\
 K2  &   29& $<15$ & -- & $<52$ & 3.16 & 0.31 & 31 & 4 & 0.47\\
 K1  &   25& $<15$ & -- & $<60$ & 4.81 & 0.23 & 28 & 2 & 0.23\\ \hline 
\multicolumn{10}{c}{1334$-$127$^*$}\\
 C   & 3822&  67.0 & 83 & 1.8  & --   & --   & --     & -- & 0.14\\ 
 K3  &  201&  37.7 & 54 & 18.8 & 2.25 & 0.06 & 147    &  1 & 0.20\\ 
 K2  &   56& $<11$ & -- & $<20$& 2.26 & 0.14 & 110    &  9 & 0.01\\ 
 K1  &   26& $<11$ & -- & $<42$& 7.18 & 0.23 & 120    &  3 & 0.20\\ \hline
\multicolumn{10}{c}{1732+389}\\
 C   & 923 & 18.7 &  5 & 2.0   & --   & --   & --   & -- & 0.32\\
 K3  & 116 &  8.0 & 33 & 6.8   & 0.80 & 0.07 & 103  &  7 & 0.08\\
 K2  &  21 & $<4$ & -- & $<19$ & 2.48 & 0.30 & 112  &  3 & 0.15\\
 K1  &  10 & $<4$ & -- & $<40$ & 3.67 & 0.19 & 108  &  5 & 0.66\\ \hline
\multicolumn{10}{c}{2131$-$021$^{\dagger}$}\\
 C  & 1011 & 18.8--22.5 & $-(28-13)$ & 1.8--2.2 & -- & -- & -- & -- & 0.11   \\
 K5 &  344 & $<6$   & --  & $<1.7$ & 0.41 & 0.03& 100 &  6 & 0.13   \\
 K4 &   19 & $<6$ & --  & $<32$ & 1.77 & 0.30& 70  & 25 & 0.14   \\
 K3 &   18 & $<6$ & --  & $<33$ & 2.82 & 0.09& 75  &  6 & 0.14   \\
 K2 &   12 & $<6$ & --  & $<50$ & 5.39 & 0.09& 86  & 10 & 0.14   \\
 K1 &   17 & $<6$ & --  & $<35$ & 6.58 & 0.04& 88  &  3 & 0.12 \\ \hline 
\multicolumn{10}{c}{2155$-$152$^{\dagger}$}\\
 C   & 207  & 21.4--46.9 & 52.0  & 10.3--22.6 & -- & -- &  --  & -- & 0.16\\ 
 K   & 844  & $<8$ & --      & $<1.0$ & 1.55 & 0.04 & $-144$ &  1 & 0.28\\ 
 K   & 897  & 49.3 & $-76.2$ & 5.5    & 4.52 & 0.05 & $-161$ &  1 & 1.50\\ 
K1   & 39   & $<8$ & --      & $<20$  & 9.41 & 0.20 & $-146$ &  1 & 0.05\\ \hline
\multicolumn{10}{l}{$^{a}$ Observed $\chi$ values for all sources 
except for 0820+225 corrected using integrated}\\
\multicolumn{10}{l}{\phantom{$^a$} rotation measures in Table~\ref{tab:rm} (see text).}\\ 
\multicolumn{10}{l}{$^{*}$ The VLA and VLBI polarizations were variable during
the VLBI run; the indicated}\\  
\multicolumn{10}{l}{\phantom{$^{*}$} $p$ and $\chi$ correspond to one two-hour interval during the run (see text).}\\
\multicolumn{10}{l}{$^{\dagger}$ The VLA and VLBI polarizations were variable during the VLBI run; the indicated}\\  
\multicolumn{10}{l}{\phantom{$^{\dagger}$} $p$ and $\chi$ give the ranges covered during the VLBI run.}\\  
\end{tabular}
\end{table*}
\begin{table*}
\centering
\def\baselinestretch{1}
\addtocounter{table}{-1}
\caption[xx]{{\sc Source Models (cont'd)}}
\begin{tabular}{cccccccccc} \hline \hline
   & $I$ & $p$ & {$\chi_{0}$}$^{a}$ & $m$ & $r$ & $\Delta r$ & $\theta$ & $\Delta\theta$ & FWHM \\ 
   & (mJy) & (mJy) & (deg) & (\%) & (mas)& (mas) & (deg) & (deg) & (mas)  \\ \hline
\multicolumn{10}{c}{2150+173}\\
C    & 222  & 11.5 & $-7$ & 5.2    & --    & --   & --    & -- & 0.13 \\
K5   & 137  & 3.2  & $-23$ & 2.3    & 0.76  & 0.02 & $-96$ & 3 & 0.79 \\       
K4   &  26  & $<2$ & --    & $<8$   & 1.51  & 0.10 & $-74$ & 5 & 0.47 \\        
K3   &  35  & 4.5  & $-4$ & 12.8   & 3.03  & 0.05 & $-85$ & 3 & 1.86 \\        
K2   &  40  & $<2$ & --    & $<5$   & 5.73  & 0.03 & $-81$ & 1 & 0.85 \\        
K1   &  45  & 2.7  & 52    & 6.0    & 6.51  & 0.03 & $-83$ & 1 & 1.01 \\ \hline 
\multicolumn{10}{c}{2254+074}\\
 C  &  104 & $< 9$ &  -- & $< 9$   & --   &  --  &  --    & -- & 0.25  \\
K3  &   20 & $< 9$ &  -- & $< 45$  & 0.85 & 0.04 & $-84$  &  9 & 0.34  \\
K2  &   25 & $< 9$ &  -- & $< 36$  & 2.70 & 0.06 & $-110$ &  3 & 1.34  \\
K1  &    7 & $< 9$ &  -- & $< 100$ & 3.69 & 0.30 & $-127$ &  6 & 0.55\\ \hline\hline
\multicolumn{10}{l}{$^{a}$ Observed $\chi$ values for all sources 
except for 0820+225 corrected using integrated}\\
\multicolumn{10}{l}{\phantom{$^a$} rotation measures in Table~\ref{tab:rm} (see text).}\\ 
\end{tabular}
\end{table*}
\begin{figure}
\mbox{
\rotate[r]{\psfig{file=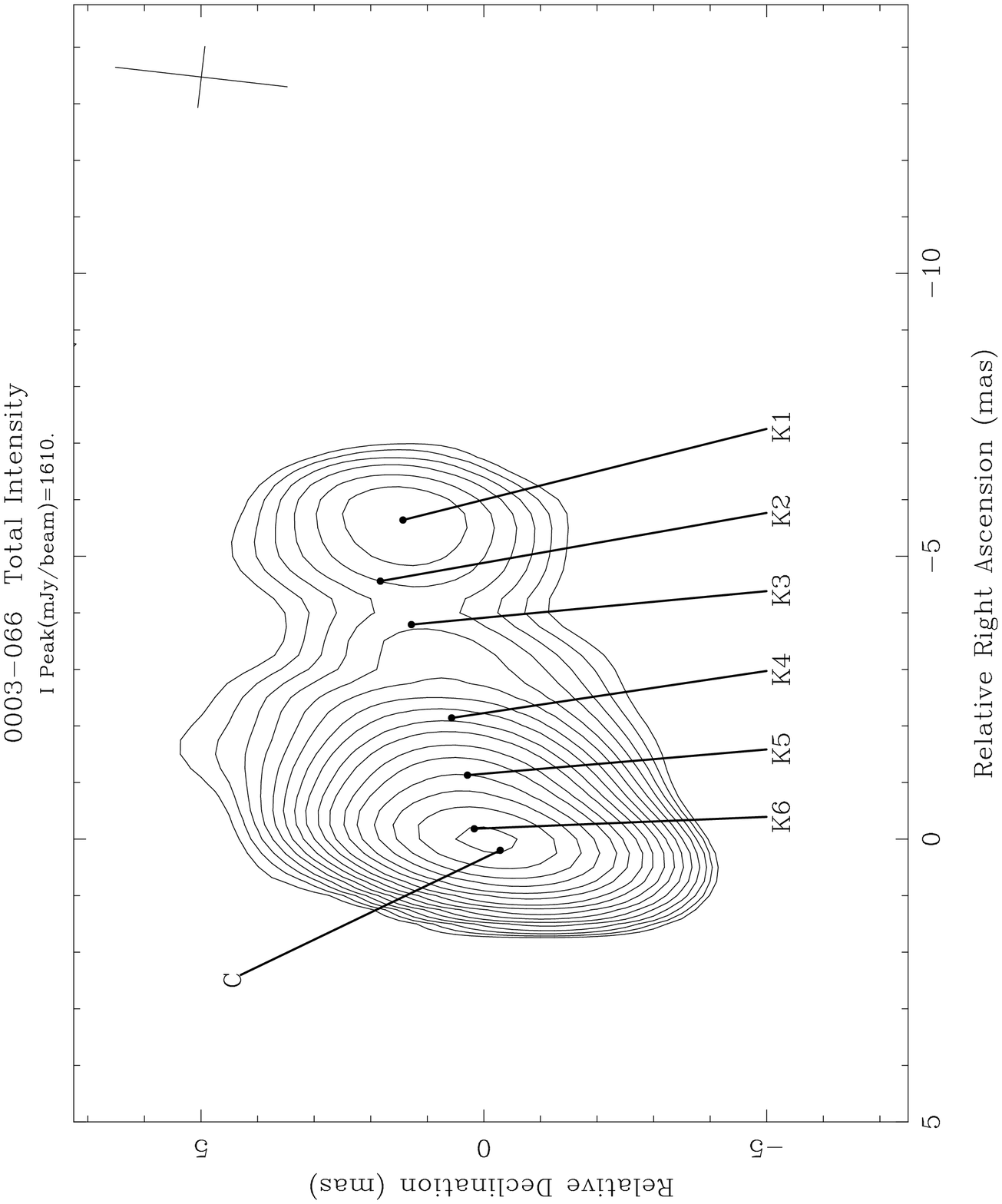,height=9.5cm}}
}
\mbox{
\rotate[r]{\psfig{file=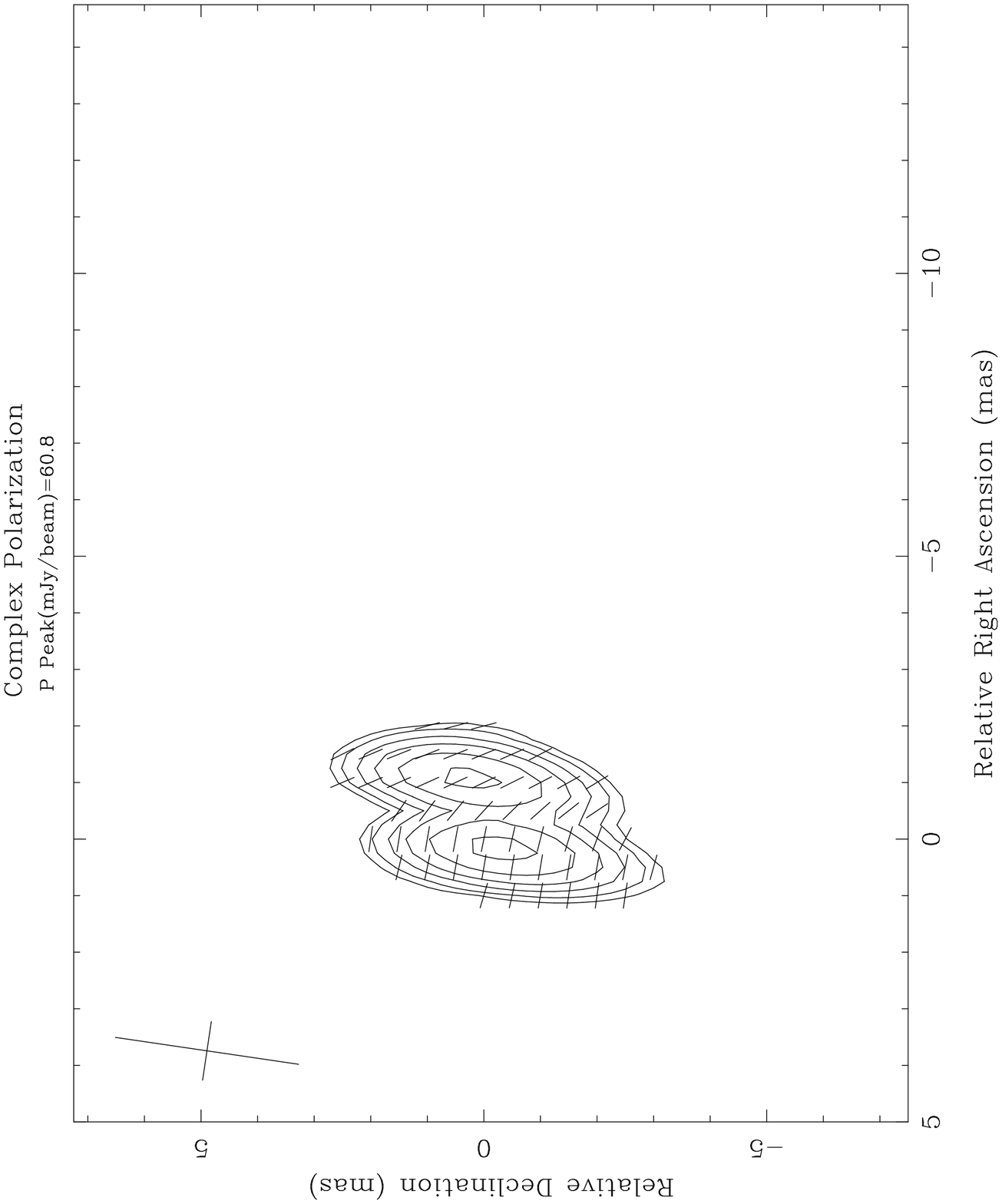,height=9.5cm}}
}
\mbox{
\rotate[r]{\psfig{file=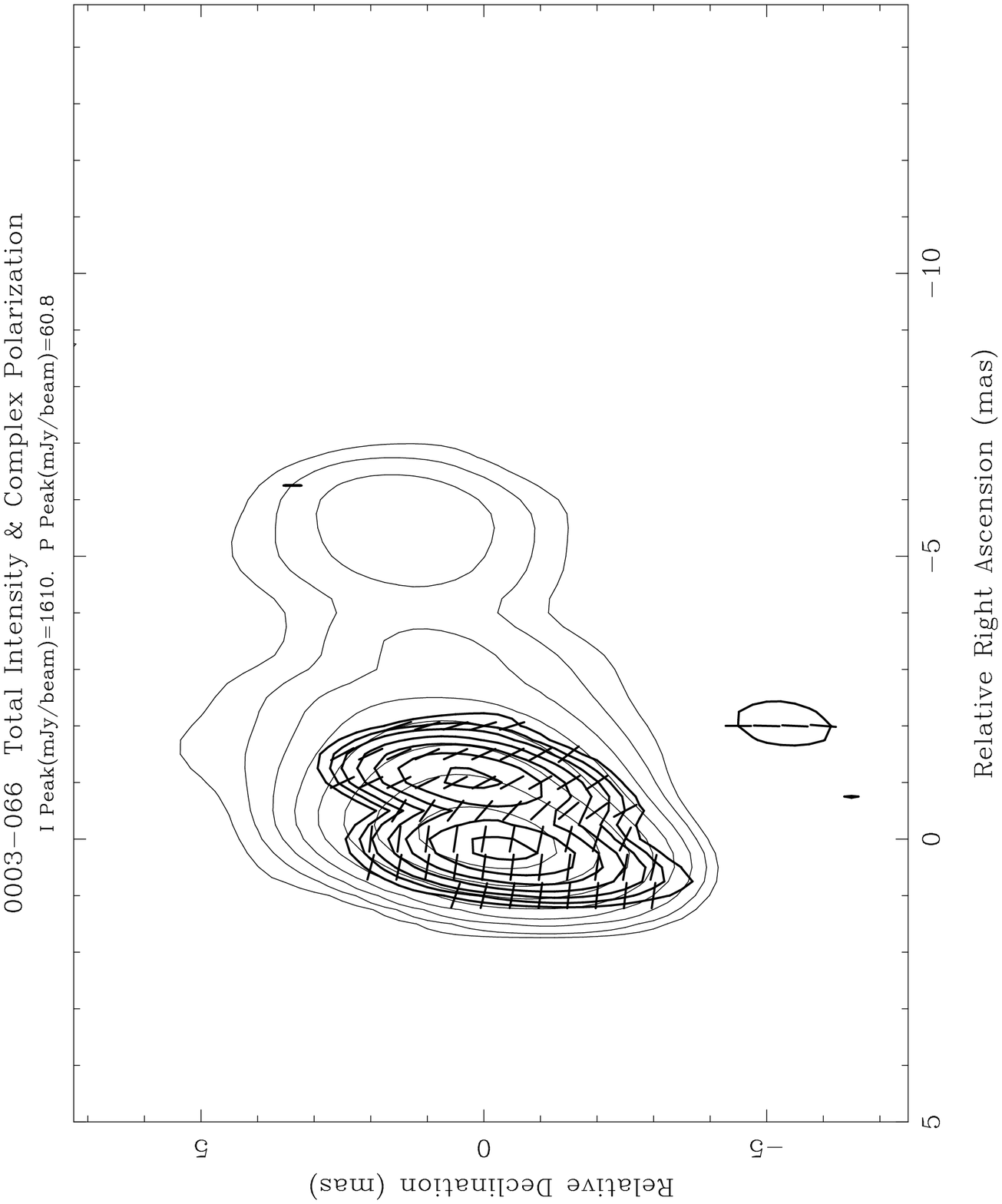,height=9.5cm}}
}
\caption{VLBI hybrid maps of 0003$-$066: (a) Total intensity, with contours 
at $-0.5$, 0.5,0.7, 1.0, 1.4, 2.0, 2.8, 4.0, 5.6, 8.0, 11, 16, 23,
32, 45, 64, and 90\% of the peak brightness of 1.61 Jy/beam. (b) Linear
polarization, with contours of polarized intensity at 16, 23, 32, 
45, 64, and 90\% of the peak brightness of 61 mJy/beam, and $\chi$ 
vectors superimposed. (c) Superposition of $P$ image (heavy lines) over
the $I$ image, with every other $I$ contour omitted.} 
\label{fig:0003}
\end{figure}
Results for the ten sources are discussed below. 
In each of the images, the restoring 
beams are shown as crosses in a corner of the images. 
For the linear 
polarization maps, the contours are those of polarized intensity $p$, and the 
plane of the electric vector is indicated by the polarization position angle 
vectors that are superimposed. 

There is always the possibility that the observed polarization position angles 
include some contribution due to Faraday rotation along the line of sight to 
the emission region. Multi-frequency VLBI studies of a number of quasars have
revealed the presence of non-uniform rotation-measure distributions on
parsec scales in some sources (Taylor 1998, 2000; Nan \etal 1999). At the
same time, other sources show no clear evidence for substantial local 
(non-Galactic) rotation measures (Gabuzda \& G\'omez 2000).  In the absence 
of simultaneous multi-wavelength VLBI polarimetry, the only Faraday 
correction that may be applied is one based
on integrated measurements. For reference, the rotation measures for the 
ten sources considered here are summarized in Table~\ref{tab:rm}.
All the rotations produced at $\lambda = 6$cm by the integrated 
rotation measures in Table~\ref{tab:rm} are less than about $15^{\circ}$; 
all except that for 1732+389 and 0820+225 are less than $10^{\circ}$ 
(Table~\ref{tab:rm}).
Therefore, application of these rotations is probably unlikely to 
introduce large errors in the resulting $\chi$ values. 
The $\chi$ values for all sources except for 0820+225 have accordingly been 
corrected using the rotation measures in Table~\ref{tab:rm}; the possibility
of offsets in $\chi$ due to local rotation measures that have not been
accounted for should be borne in mind, however. The case of
0820+225 is discussed below: for this source, quasi-simutaneous 
multi-frequency VLBA polarization images (Gabuzda, Pushkarev \& Garnich,
in prep.) have revealed a non-uniform
rotation-measure distribution on milliarcsecond scales, so that it is
clearly not appropriate to apply the integrated rotation measure to each
of the VLBI components.  

\begin{figure}
\mbox{
\rotate[r]{\psfig{file=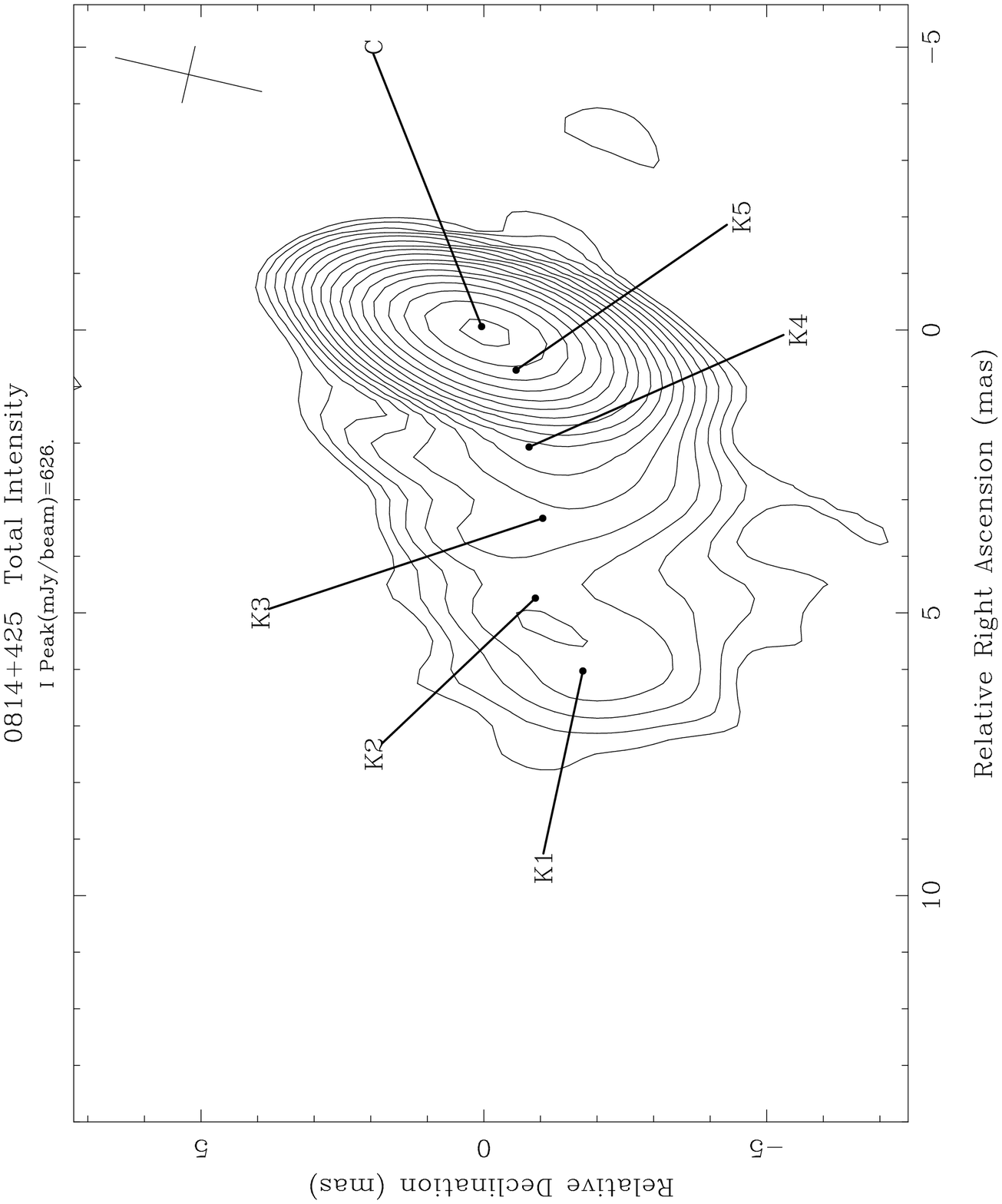,height=9.5cm}}
}
\mbox{
\rotate[r]{\psfig{file=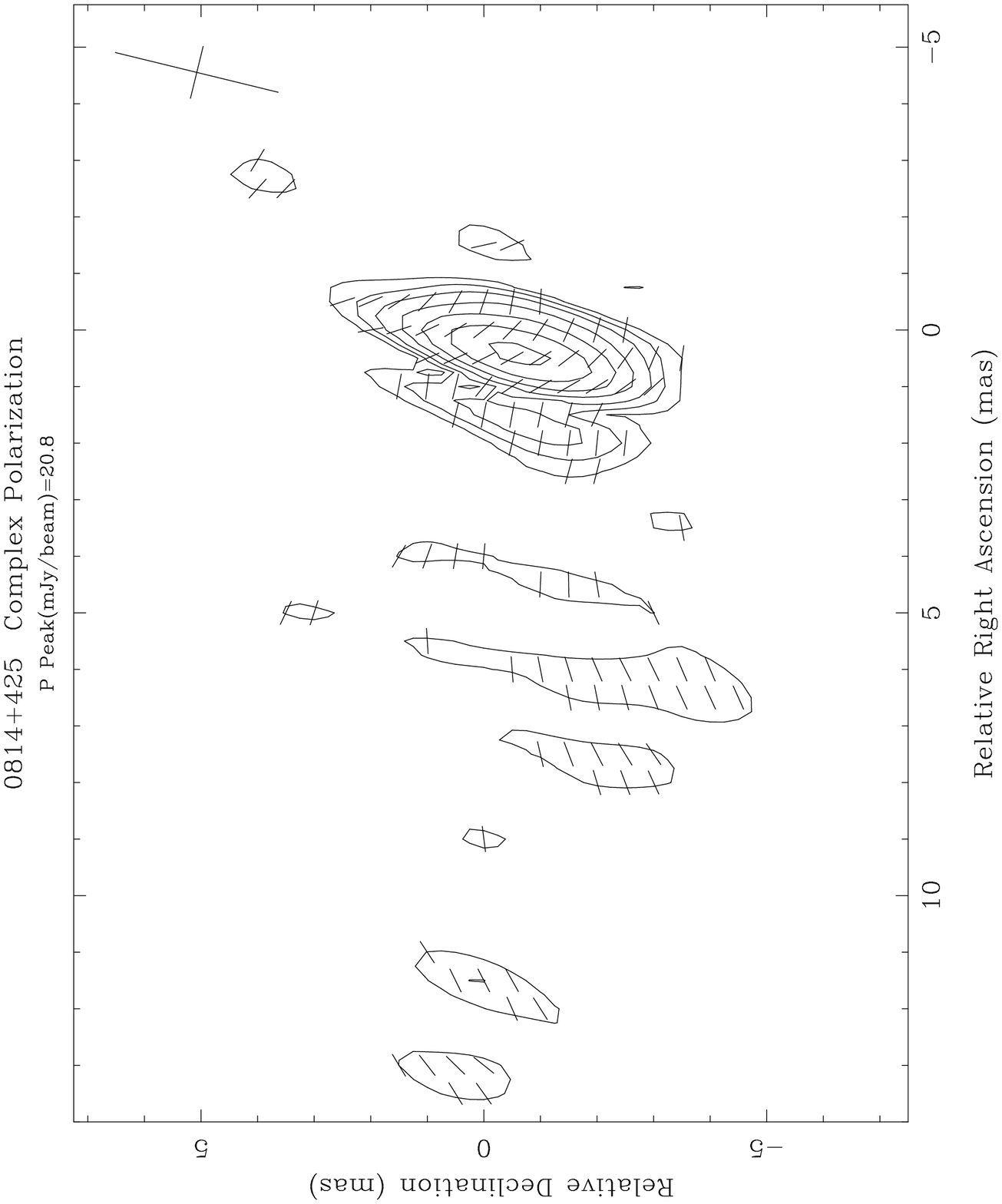,height=9.5cm}}
}
\mbox{
\rotate[r]{\psfig{file=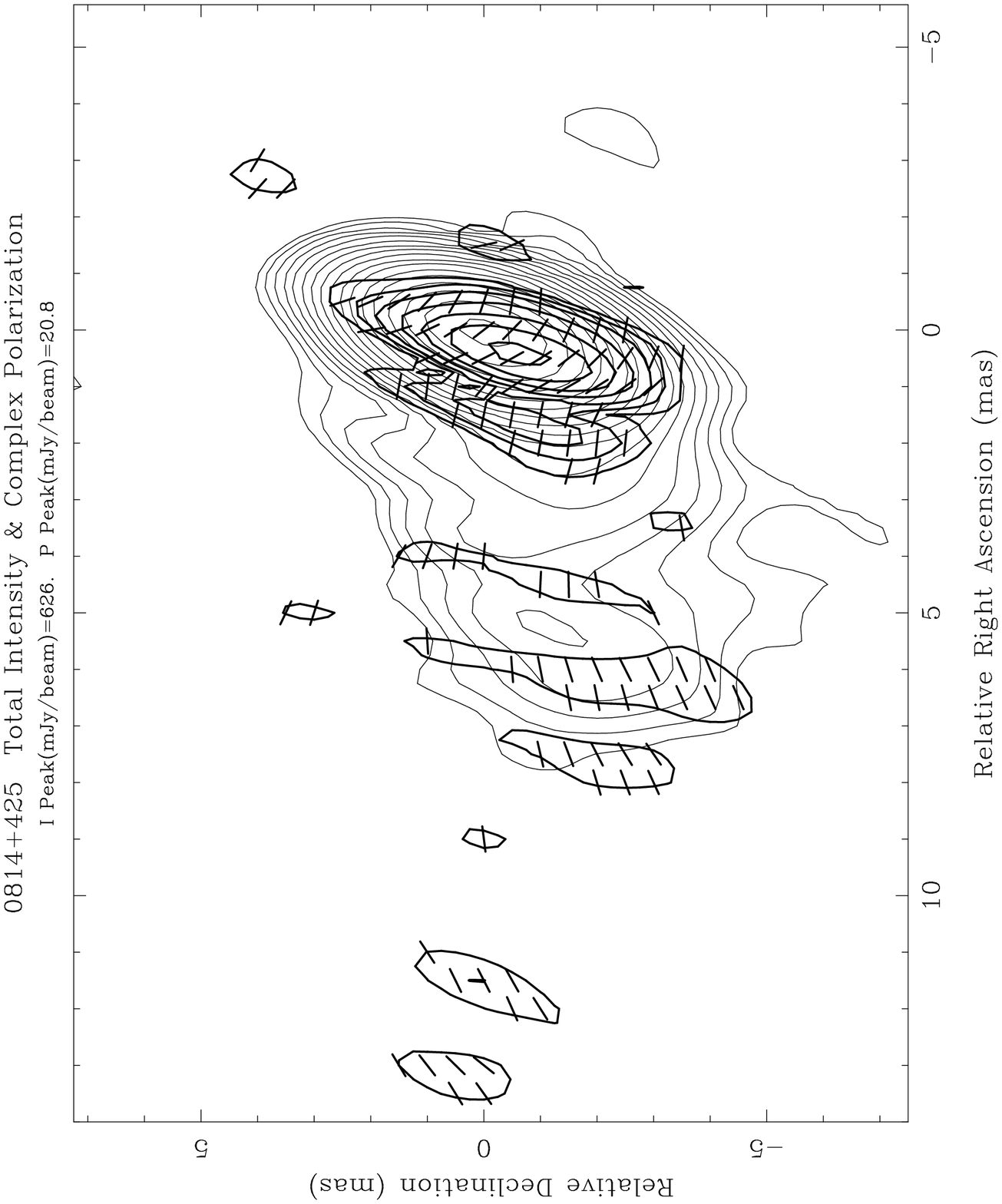,height=9.5cm}}
}
\caption{Total intensity VLBI hybrid map of 0814+425, with contours 
at $-0.3$, 0.3, 0.5, 0.7, 1.0, 1.4, 2.0, 2.8, 4.0, 5.6, 8.0, 11, 16, 23,
32, 45, 64, and 90\% of the peak brightness of 0.63 Jy/beam. (b) Linear
polarization, with contours of polarized intensity at 11, 16, 23, 32,
45, 64, and 90\% of the peak brightness of 21 mJy/beam, and $\chi$
vectors superimposed. (c) Superposition of $P$ image (heavy lines) over
the $I$ image, with every other $I$ contour omitted.} 
\label{fig:0814}
\end{figure}

\begin{figure}
\mbox{
\rotate[r]{\psfig{file=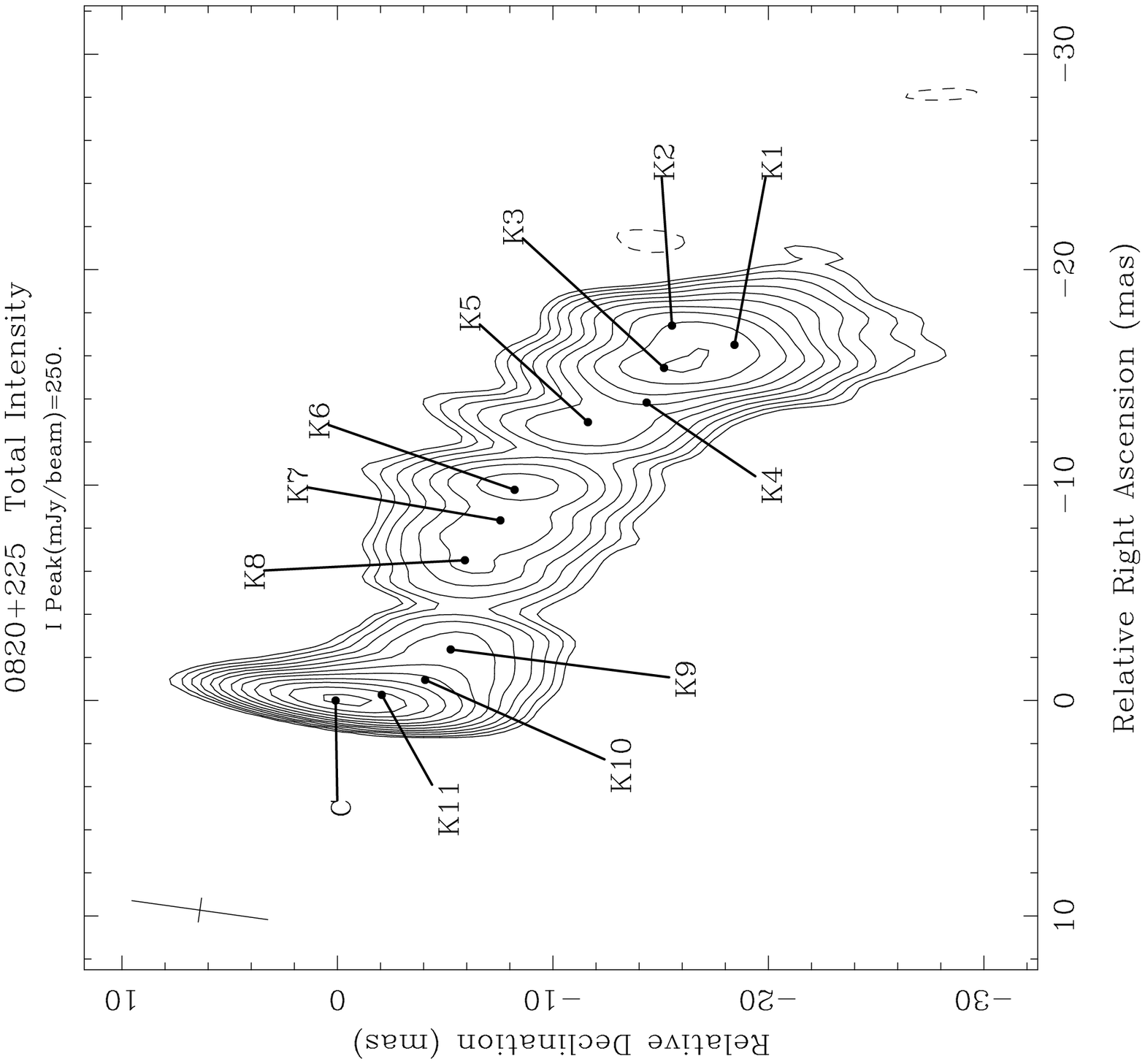,height=9.5cm}}
}
\mbox{
\rotate[r]{\psfig{file=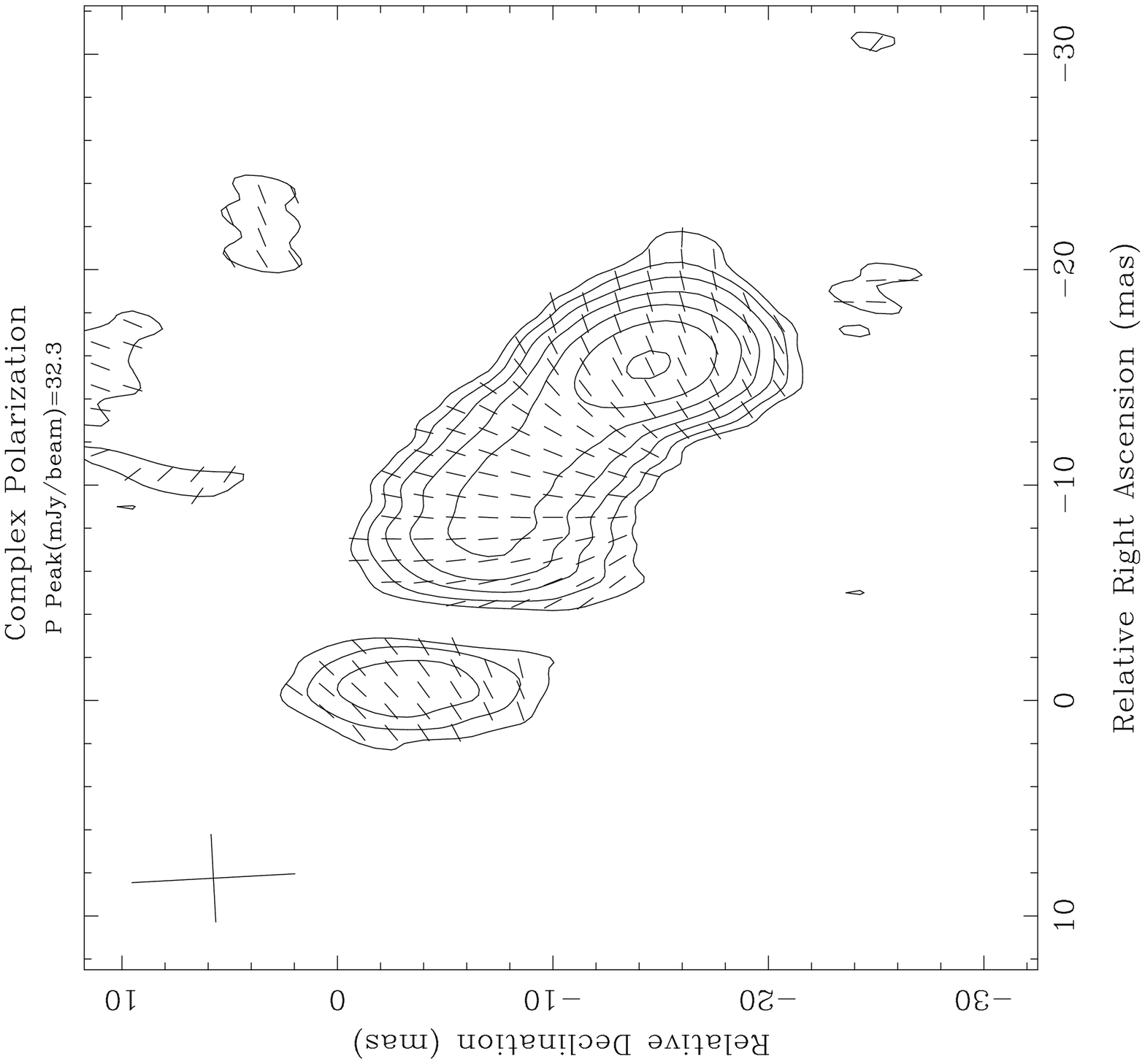,height=9.5cm}}
}
\mbox{
\rotate[r]{\psfig{file=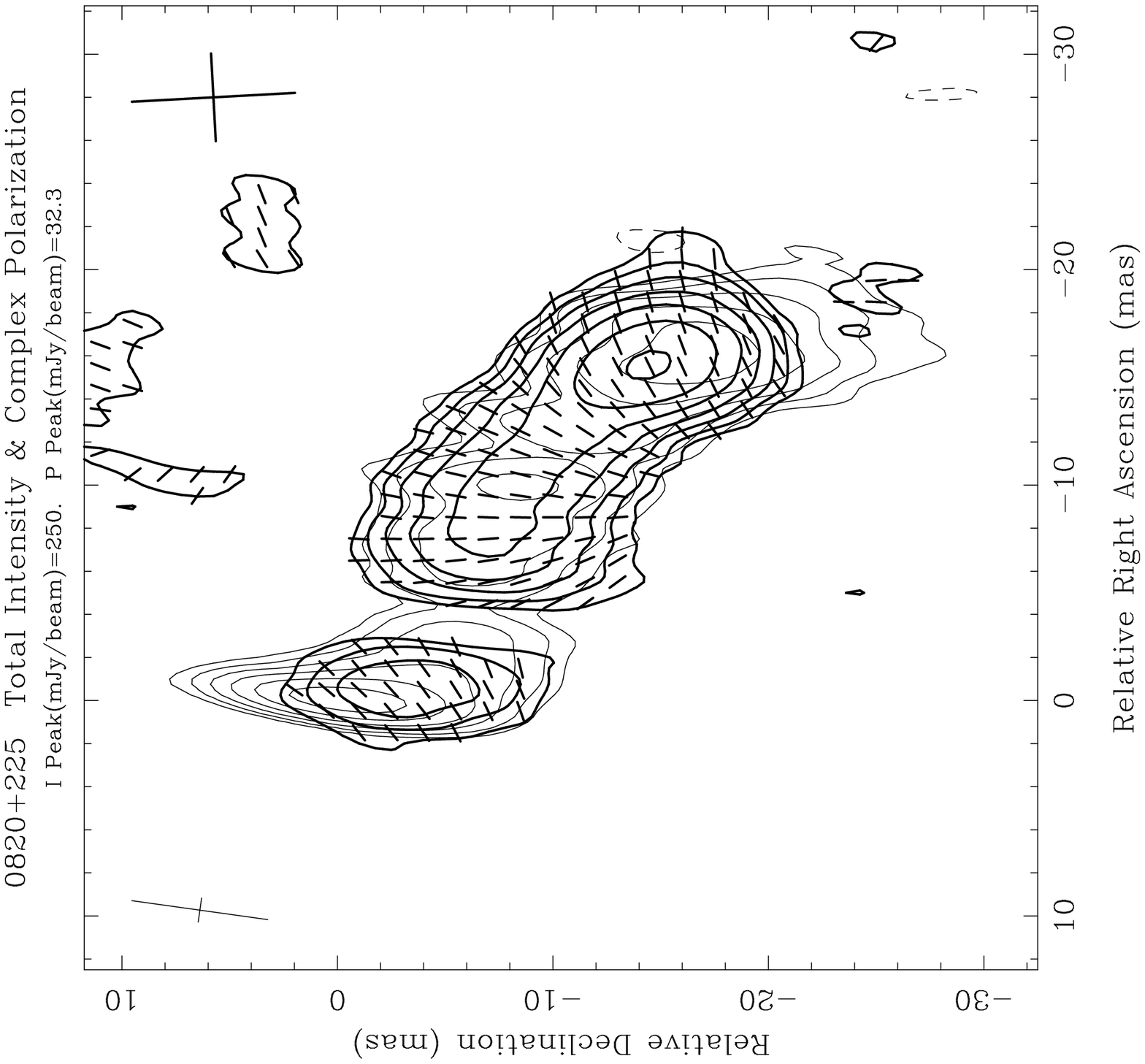,height=9.5cm}}
}
\caption{Total intensity VLBI hybrid map of 0820+225, with contours 
at $-2.0$, 2.0, 2.8, 4.0, 5.6, 8.0, 11, 16, 23,
32, 45, 64, and 90\% of the peak brightness of 0.25 Jy/beam. (b) Linear
polarization, with contours of polarized intensity at 12, 17, 24, 34, 
48, 68, and 96\% of the peak brightness of 32 mJy/beam, and $\chi$
vectors superimposed. (c) Superposition of $P$ image (heavy lines) over
the $I$ image, with every other $I$ contour omitted.} 
\label{fig:0820}
\end{figure}

Models for the source structures were derived by fitting the complex $I$ and 
$P$ visibilities that come from the hybrid mapping process as described by 
Roberts, Gabuzda and Wardle (1987) and Gabuzda, Wardle and Roberts (1989b). The 
model fits are shown in Table~\ref{tab:models}. We denote distance from the
core by $r$ and the VLBI jet direction by $\theta$. In all cases, we
checked for consistency between the model-fitting results, the distribution
of CLEAN components, and the visual appearance of the images. We included 
components in the models only if there was clear evidence for them in the
CLEAN components derived during the imaging process. 

The $I$ and $P$ 
visibilities were fit separately, in order to allow for small differences
in the positions and sizes of corresponding $I$ and $P$ components, either
intrinsic to the source structure or associated with residual calibration
errors. When $P$ components have been identified with specific $I$ 
components, the two positions agree to within their 3$\sigma$ errors. 
When calculating the degrees of polarization $m$ for individual features, 
we have not taken into account possible small physical offsets between the
corresponding $I$ and $P$ components, so that the $m$ values in 
Table~\ref{tab:models} represent averages for each component.

The errors 
of the separations of jet 
components from the core given in Table~\ref{tab:models} are formal 1$\sigma$ 
errors, corresponding to an increase in the best-fit $\chi^{2}$ by unity. The 
smallest of these formal errors almost certainly underestimate the actual 
errors; realistically, the smallest 1$\sigma$ errors in component separations 
are probably no less than $\sim 0.05$ mas.

In our 1992.23 VLBI run, the polarizations of three of the eighteen 
sources---1334--127, 2131--021, and 2155--152---varied during the VLBI 
observations. We were able to detect these
variations since the phased VLA was included in the VLB
array, so that we had measurements of the integrated total intensity and
polarization during each of the VLBI scans. Summary information about the
polarizations of these sources is given below; the polarization
variations are analysed in detail in an accompanying paper by 
Gabuzda \etal (2000). 

We assume throughout a 
Friedmann universe with Hubble constant of $100 h$\,km\,sec$^{-1}$\,Mpc$^{-1}$ 
and $q_{0}=0.5$. 

\begin{figure}
\mbox{
\rotate[r]{\psfig{file=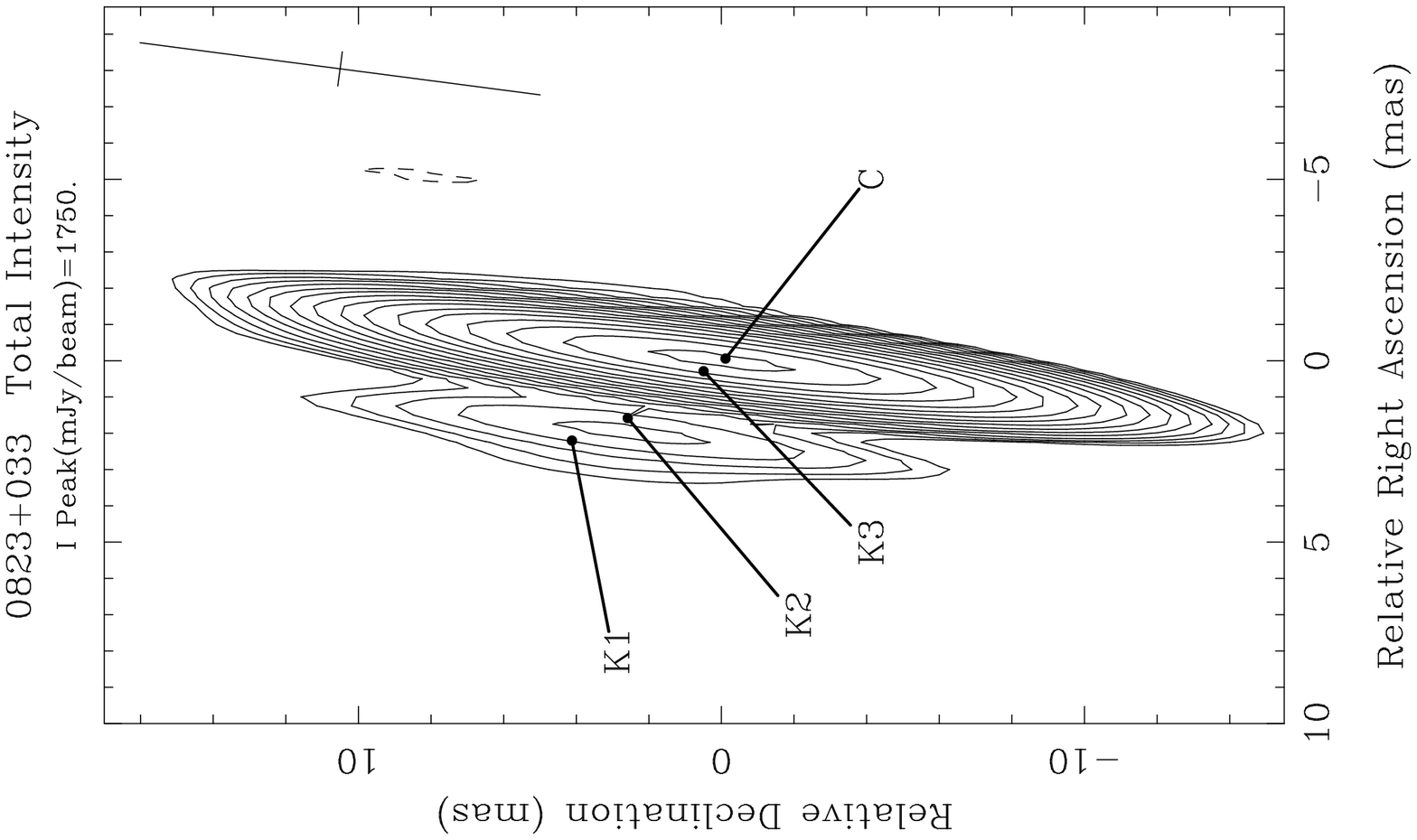,height=9.5cm}}
}
\mbox{
\rotate[r]{\psfig{file=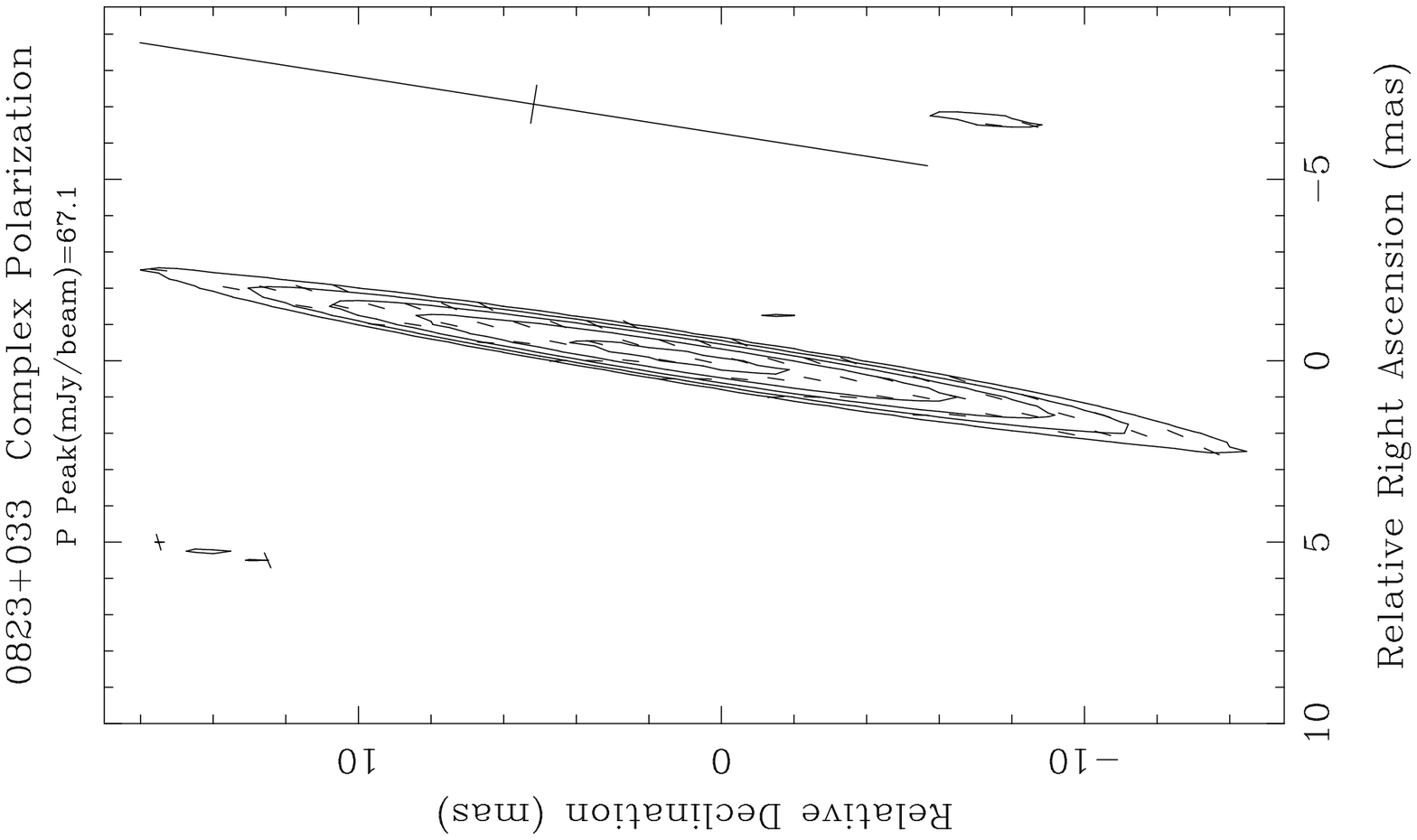,height=9.5cm}}
}
\mbox{
\rotate[r]{\psfig{file=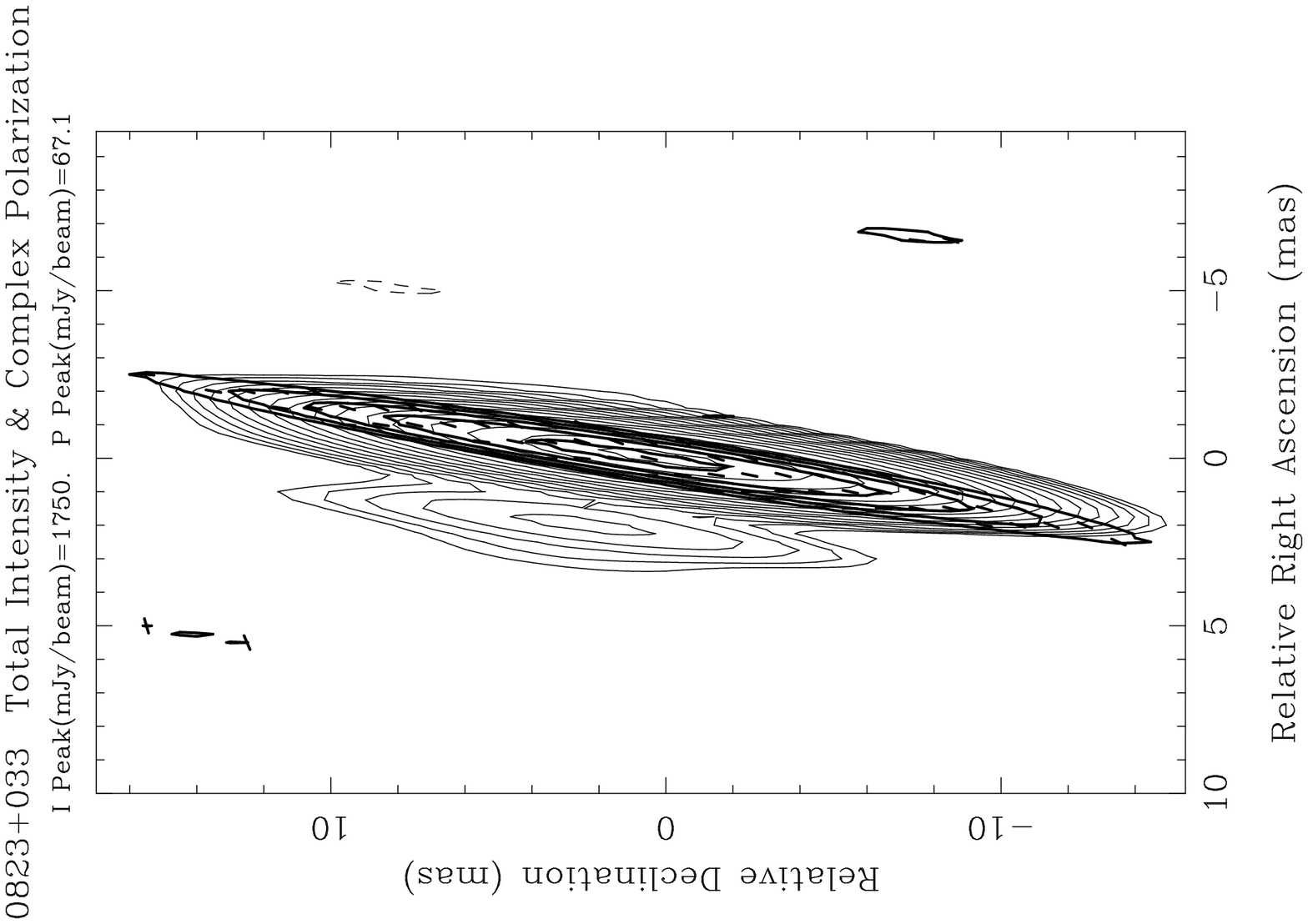,height=9.5cm}}
}
\caption{Total intensity VLBI hybrid map of $0823+033$, with contours 
at $-0.5$, 0.5, 0.7, 1.0, 1.4, 2.0, 2.8, 4.0, 5.6, 8.0, 11, 16, 23,
32, 45, 64, and 90\% of the peak brightness of 1.75 Jy/beam. (b) Linear
polarization, with contours of polarized intensity at 23, 32,
45, 64, and 90\% of the peak brightness of 67 mJy/beam, and $\chi$
vectors superimposed. (c) Superposition of $P$ image (heavy lines) over
the $I$ image, with every other $I$ contour omitted.} 
\label{fig:0823}
\end{figure}
 
\begin{figure}
\mbox{
\rotate[r]{\psfig{file=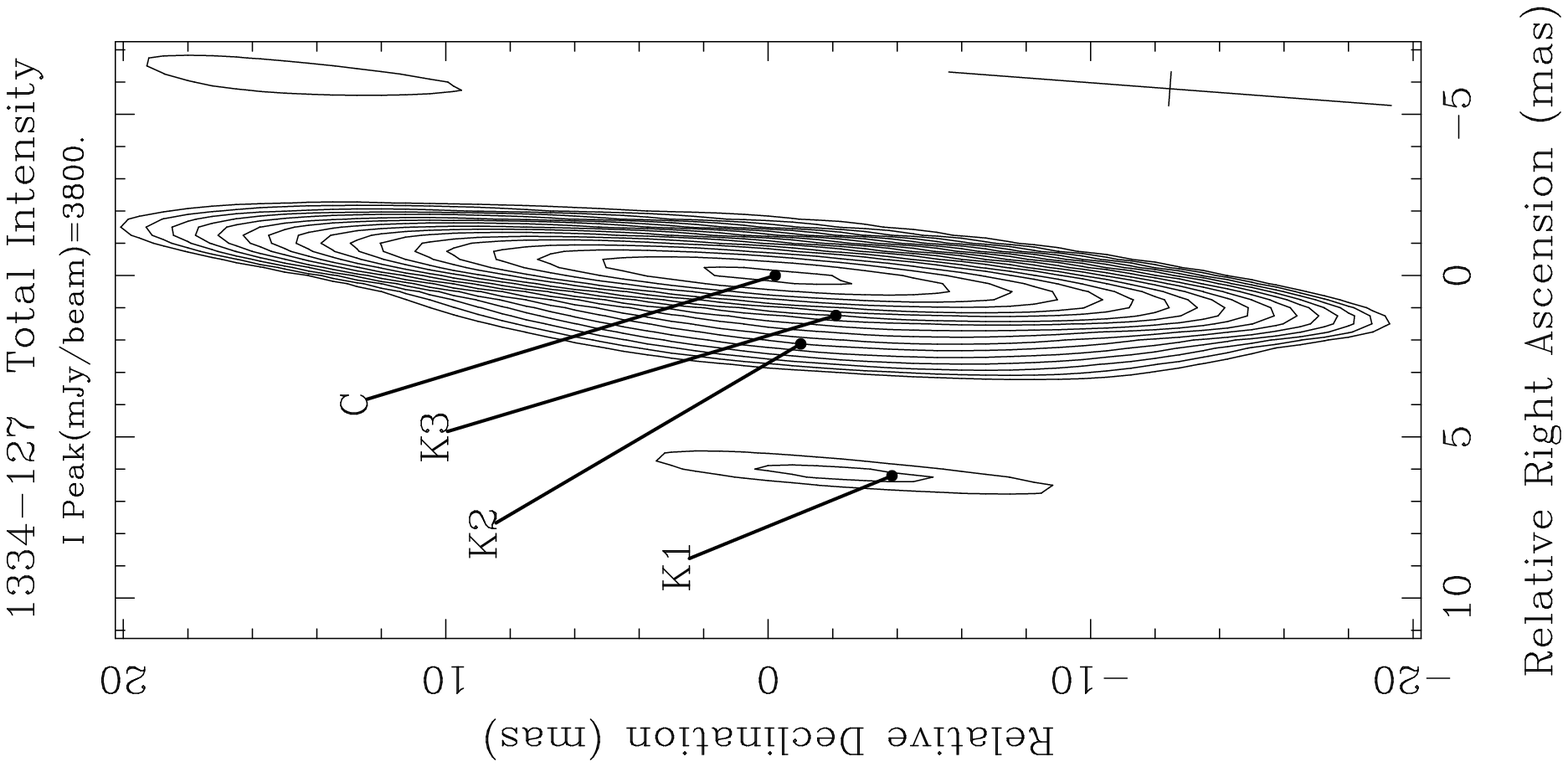,height=9.5cm}}
}
\mbox{
\rotate[r]{\psfig{file=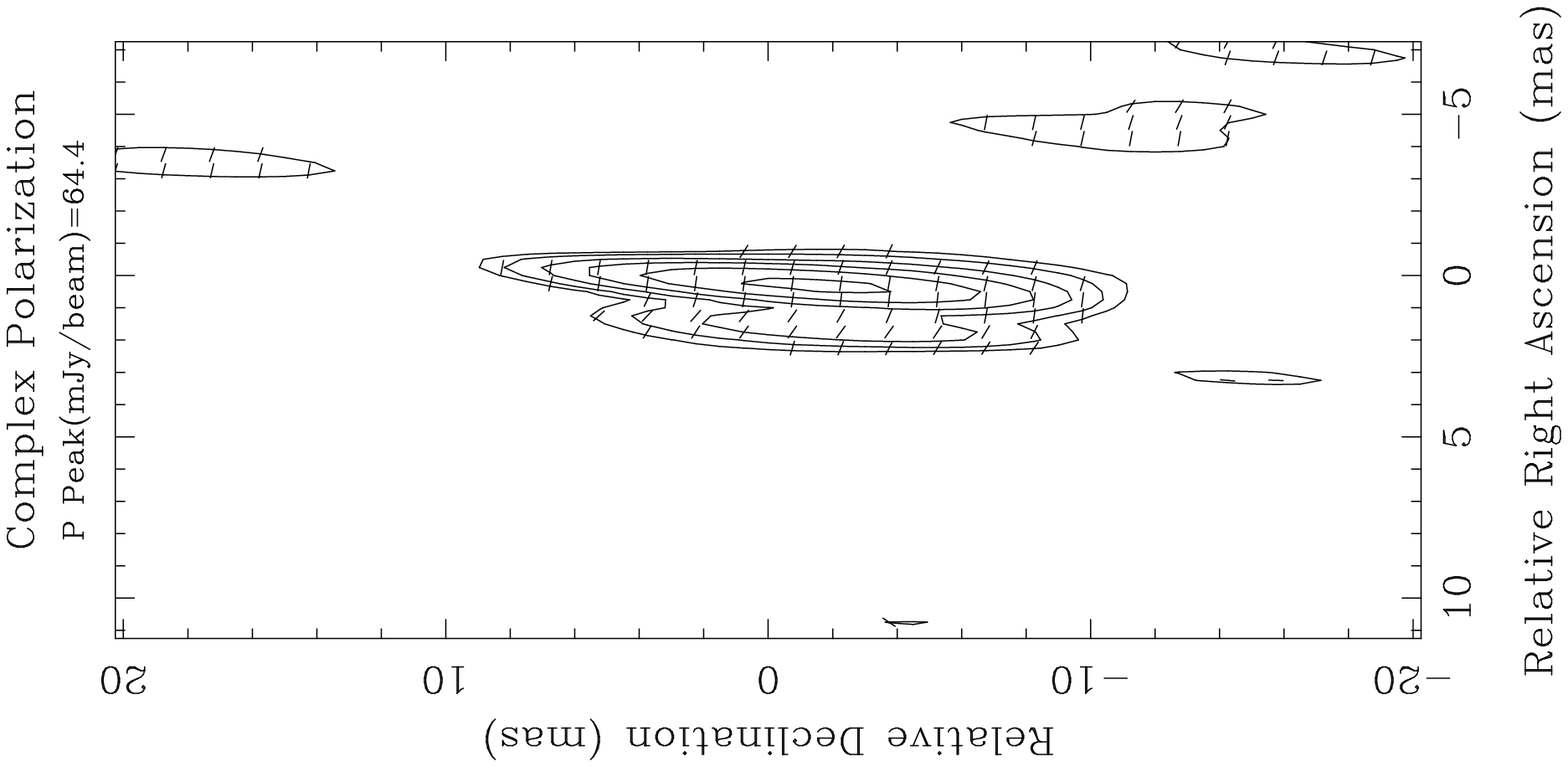,height=9.5cm}}
}
\mbox{
\rotate[r]{\psfig{file=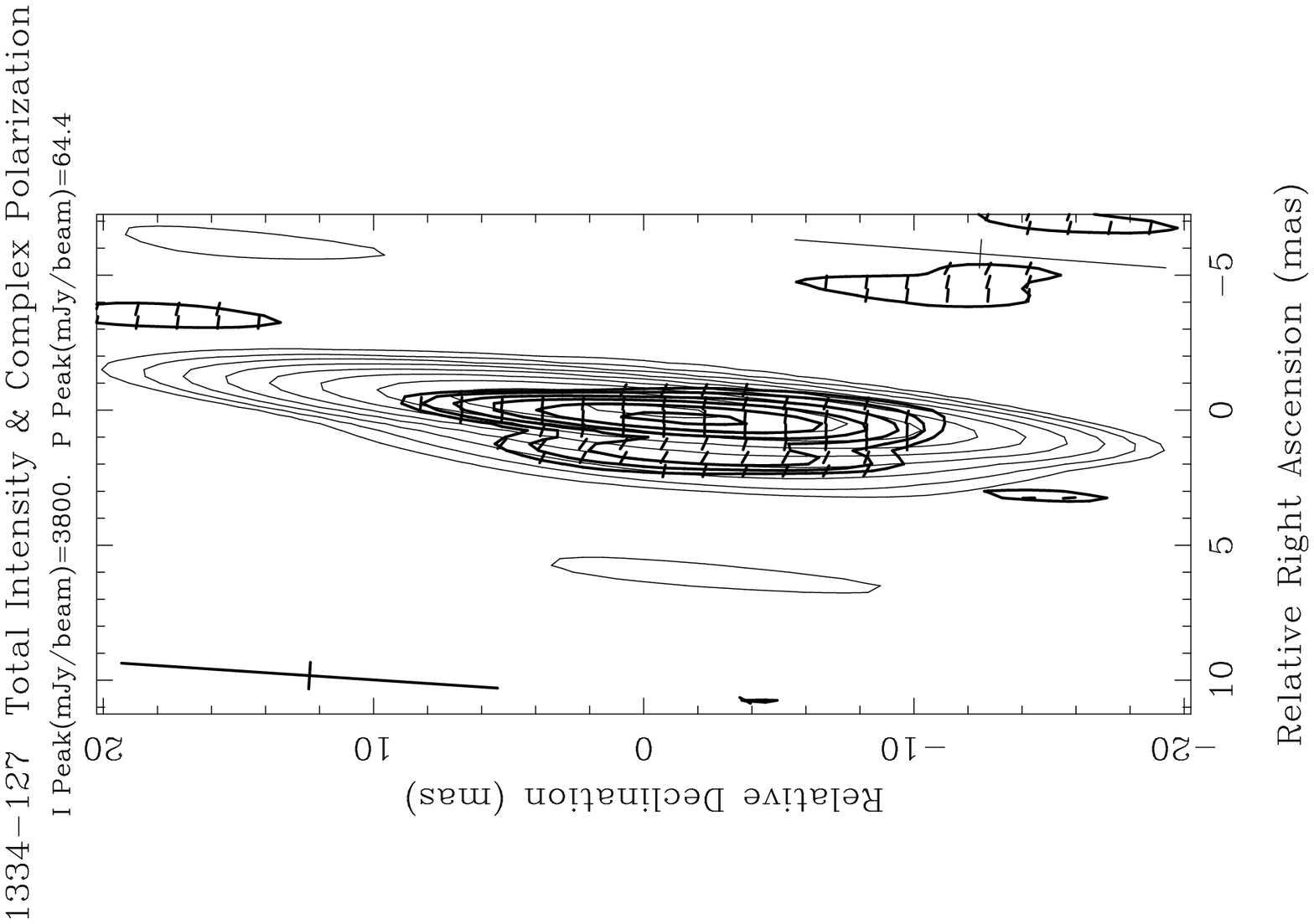,height=9.5cm}}
}
\caption{Total intensity VLBI hybrid map of $1334-127$, with contours 
at $-0.35$, 0.5, 0.7, 1.0, 1.4, 2.0, 2.8, 4.0, 5.6, 8.0, 11, 16, 23,
32, 45, 64, and 90\% of the peak brightness of 3.80 Jy/beam. (b) Linear
polarization, with contours of polarized intensity at 17, 24, 34,
48, 68, and 96\% of the peak brightness of 64 mJy/beam, and $\chi$
vectors superimposed. (c) Superposition of $P$ image (heavy lines) over
the $I$ image, with every other $I$ contour omitted.} 
\label{fig:1334}
\end{figure}
 
\subsection{0003--066}

The source 0003--066 has a redshift of $z = 0.35$. The 2~cm image of
Kellermann \etal (1998) and 13 and 4~cm images of Fey \& Charlot (1997) 
show a jet extending toward the northwest.  Our $I$ image 
(Fig.~\ref{fig:0003}) clearly shows
the straight jet in position angle $\theta = -75^{\circ}$,
in which we have modelled six components (Fig.~\ref{fig:0003}). The
linear polarization map shows a double structure, where one component
corresponds to the core and the other to the jet feature K5. 
The polarization position angle $\chi$ in the jet is nearly perpendicular
to the jet axis. 

\begin{figure}
\mbox{
\rotate[r]{\psfig{file=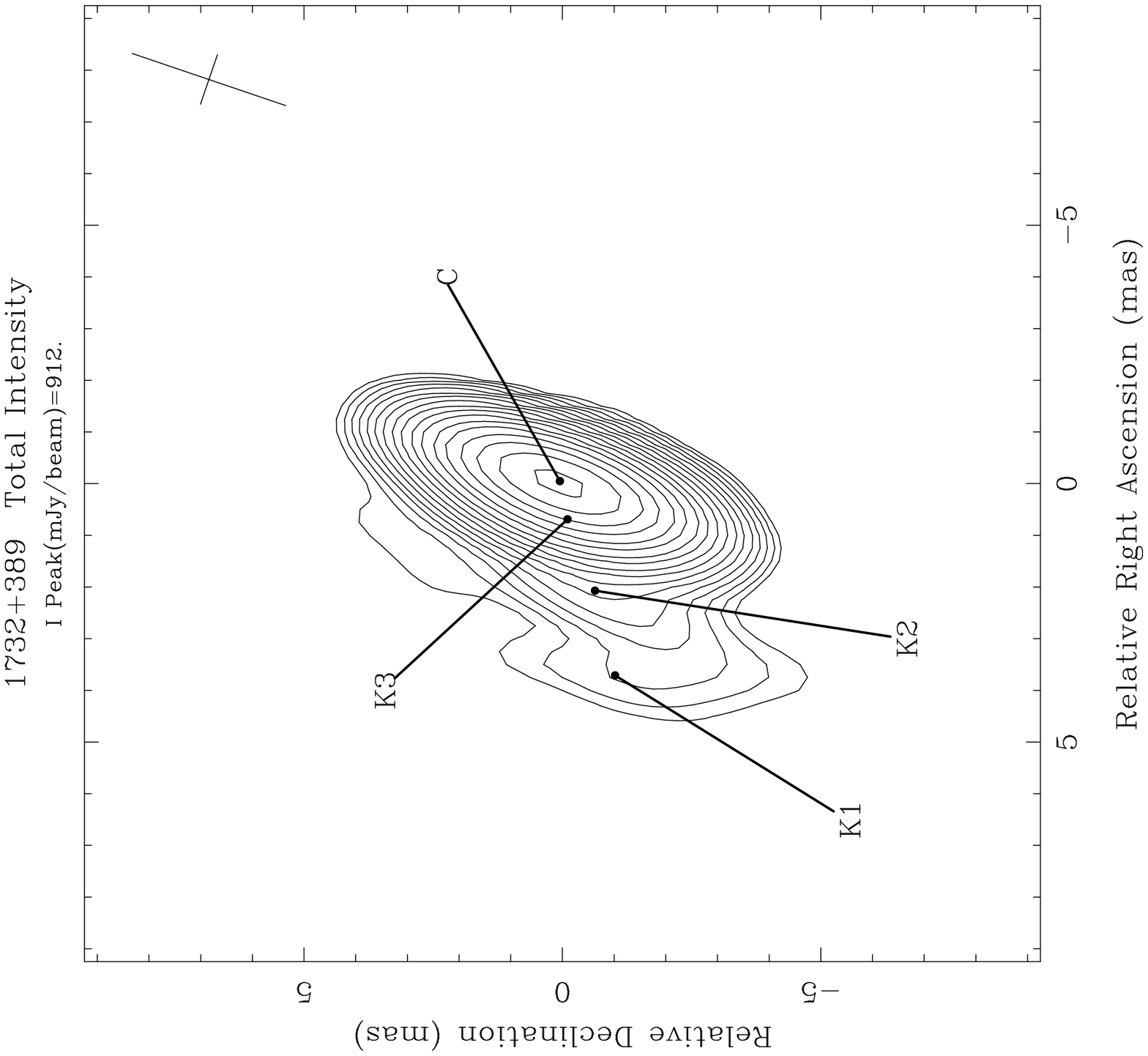,height=9.5cm}}
}
\mbox{
\rotate[r]{\psfig{file=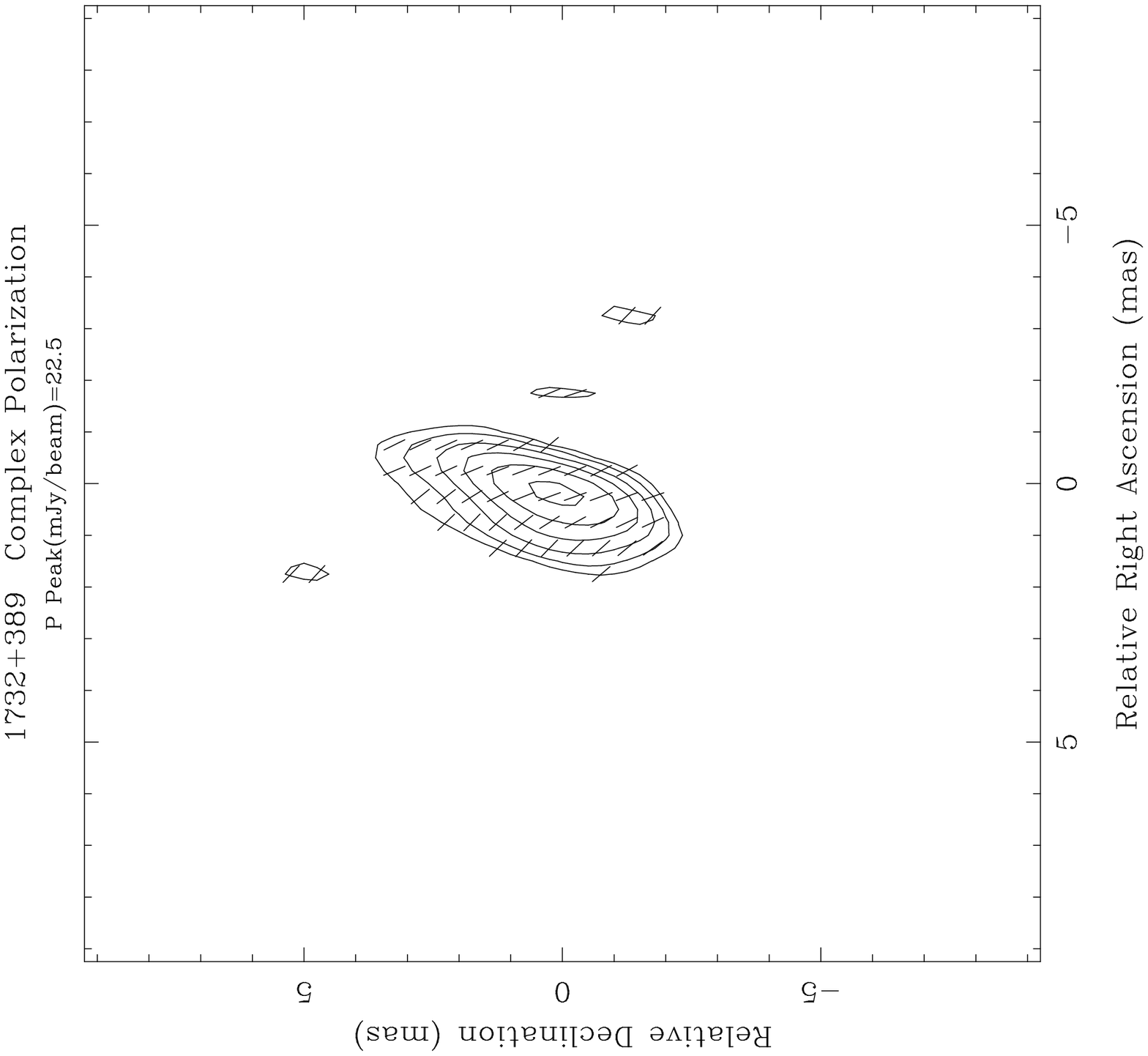,height=9.5cm}}
}
\mbox{
\rotate[r]{\psfig{file=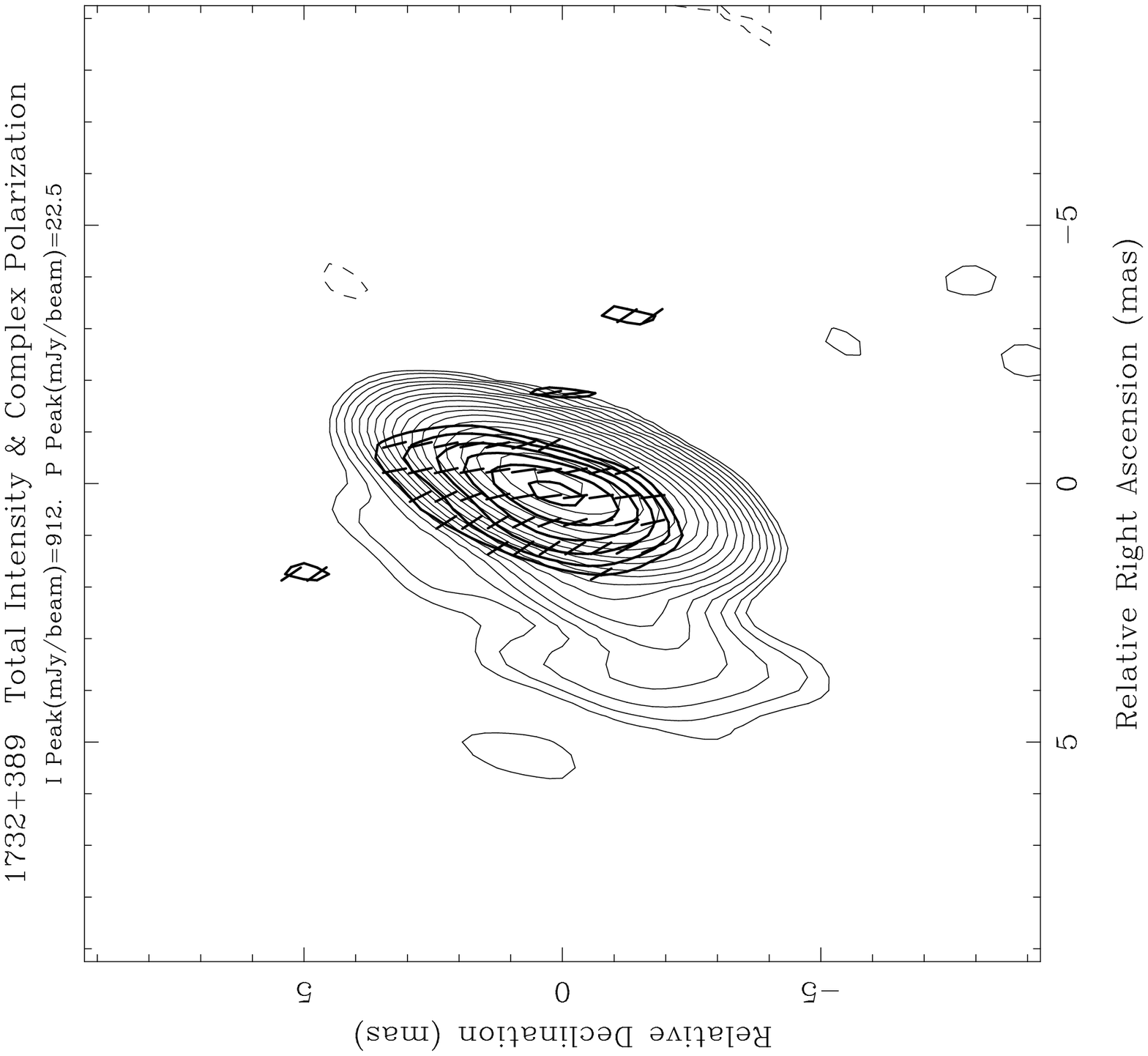,height=9.5cm}}
}
\caption{Total intensity VLBI hybrid map of 1732+389, with contours 
at $-0.35$, 0.35, 0.5, 0.7, 1.0, 1.4, 2.0, 2.8, 4.0, 5.6, 8.0, 11, 16, 23,
32, 45, 64, and 90\% of the peak brightness of 0.91 Jy/beam. (b) Linear
polarization, with contours of polarized intensity at 16, 23, 32, 
45, 64, and 90\% of the peak brightness of 22 mJy/beam, and $\chi$
vectors superimposed. (c) Superposition of $P$ image (heavy lines) over
the $I$ image, with every other $I$ contour omitted.} 
\label{fig:1732}
\end{figure}

\subsection{0814+425}

Observations by Wills \& Wills (1976) provided evidence that the redshift 
of this object is $z = 0.258$. Subsequent observations (Stickel \etal 1993a)
have not directly confirmed this value, though the CFHT observations of
Wurtz \etal (1996) showed the host galaxy to be resolved, consistent with
a redshift of 0.258.  The 2~cm image of Kellermann \etal (1998) for
epoch 1995 April shows 
a jet that emerges to the east, then appears to bend sharply to the south.  
Our 6~cm images (Fig.~\ref{fig:0814}) suggest a more gently curving jet 
structure that initially
emerges in position angle $\sim 130^{\circ}$, then turns more nearly east. 
Our images do not show clear evidence for the sharp bending observed by
Kellermann \etal (1998), even in the innermost part of the jet. 
VLBA images for epoch 1997 February at 2, 4, and 
6~cm currently being analyzed (Pushkarev et~al., in prep.) clearly show
the predominant jet direction to be nearly due east, with the 2~cm image
in good qualitative agreement with that of Kellermann \etal (1998).  
This suggests that the compact eastern extention in the 2~cm images
referred to above may have emerged after the 6~cm observations presented
here. We detect polarization in the two
innermost jet components in our images; $\chi$ appears to be transverse
to the local jet direction in K5 and aligned with the local jet direction
in K4. 

\begin{figure}
\mbox{
\rotate[r]{\psfig{file=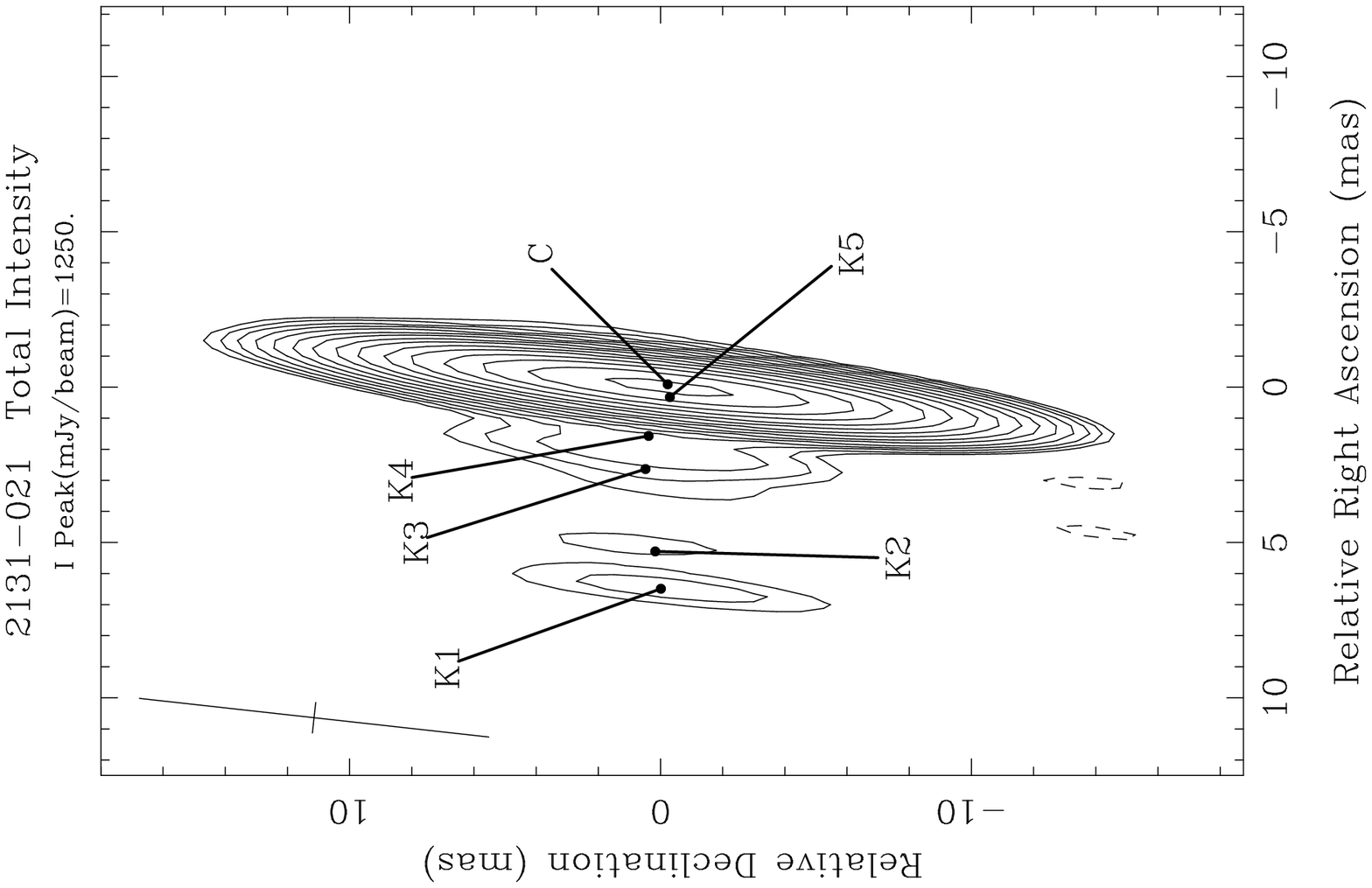,height=9.5cm}}
}
\mbox{
\rotate[r]{\psfig{file=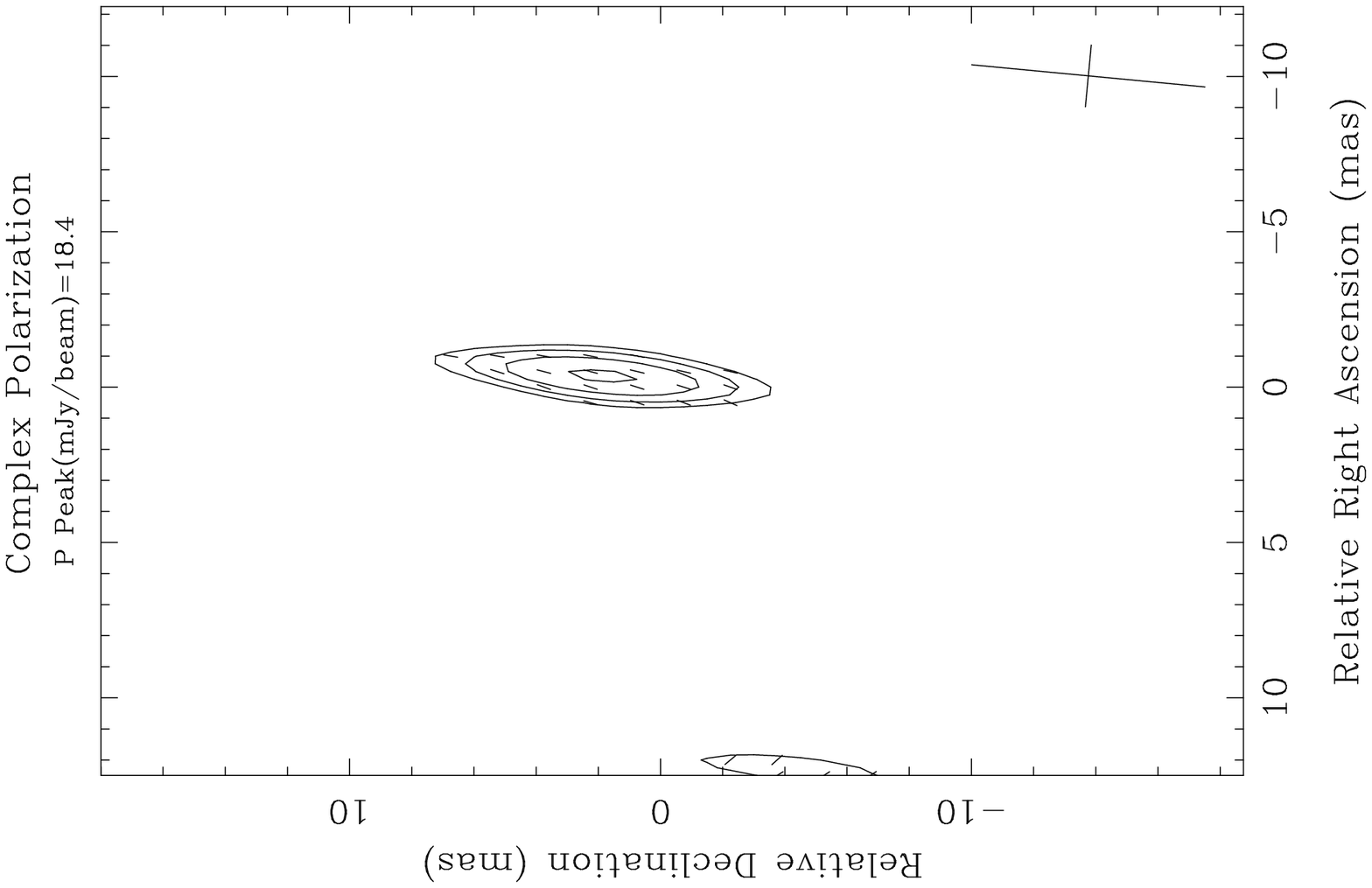,height=9.5cm}}
}
\mbox{
\rotate[r]{\psfig{file=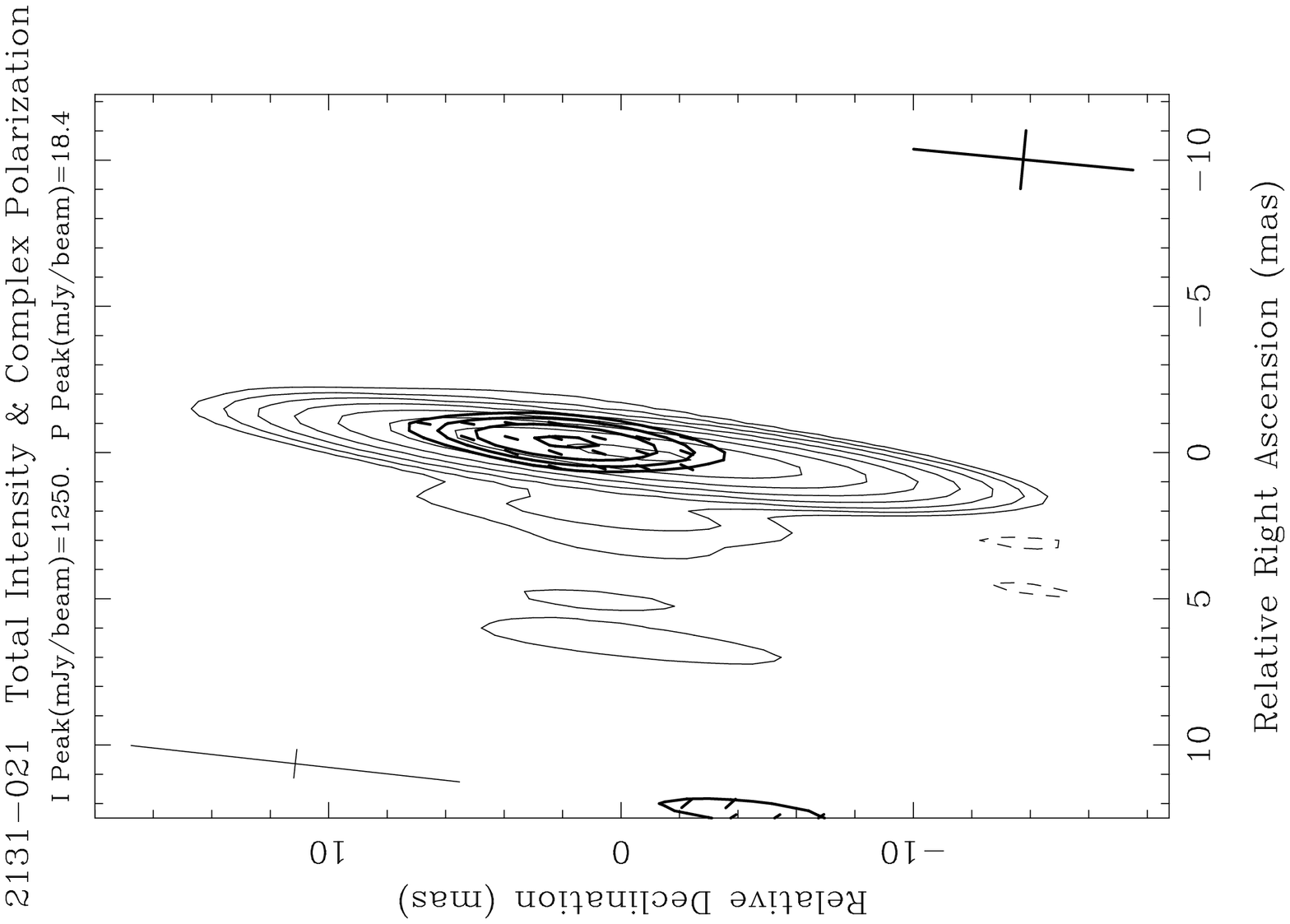,height=9.5cm}}
}
\caption{Total intensity VLBI hybrid map of $2131-021$, with contours 
at $-0.5$, 0.5, 0.7, 1.0, 1.4, 2.0, 2.8, 4.0, 5.6, 8.0, 11, 16, 23,
32, 45, 64, and 90\% of the peak brightness of 0.95 Jy/beam. (b) Linear
polarization, with contours of polarized intensity at 34, 
48, 68, and 96\% of the peak brightness of 18 mJy/beam, and $\chi$
vectors superimposed. (c) Superposition of $P$ image (heavy lines) over
the $I$ image, with every other $I$ contour omitted.} 
\label{fig:2131}
\end{figure}
 
\subsection{0820+225}

The redshift of this source is $z = 0.95$ (Stickel \etal 1993a), making
it one of the most distant objects in the sample. As far as we are aware,
ours are the first VLBI images of this source. It displays a rather
extraordinary VLBI jet, with very rich $I$ and $P$ structure.
There is much more extended emission than observed
in most of the sample objects; this is particularly striking given its
large redshift. The jet extends nearly 30~mas to the southwest, and only a
small fraction of the total flux in the VLBI image is contained
in the VLBI core. We have modelled the image as a core plus 11 jet
components. We expect that this is a good description of the source within
15--20~mas from the core, but our jet components K1--K4 represent only
an approximate model for the bright, extended region at the end of the
visible jet. The observed orientation of $\chi$ is transverse to 
the jet; however,  multi-frequency VLBI polarization observations
(Gabuzda and Pushkarev, in preparation) indicate the presence of a large
local rotation measure in the inner jet (near K10), and the derotated 
$\chi$ values in this region are well aligned with the local jet direction. 

\begin{figure}
\mbox{
\rotate[r]{\psfig{file=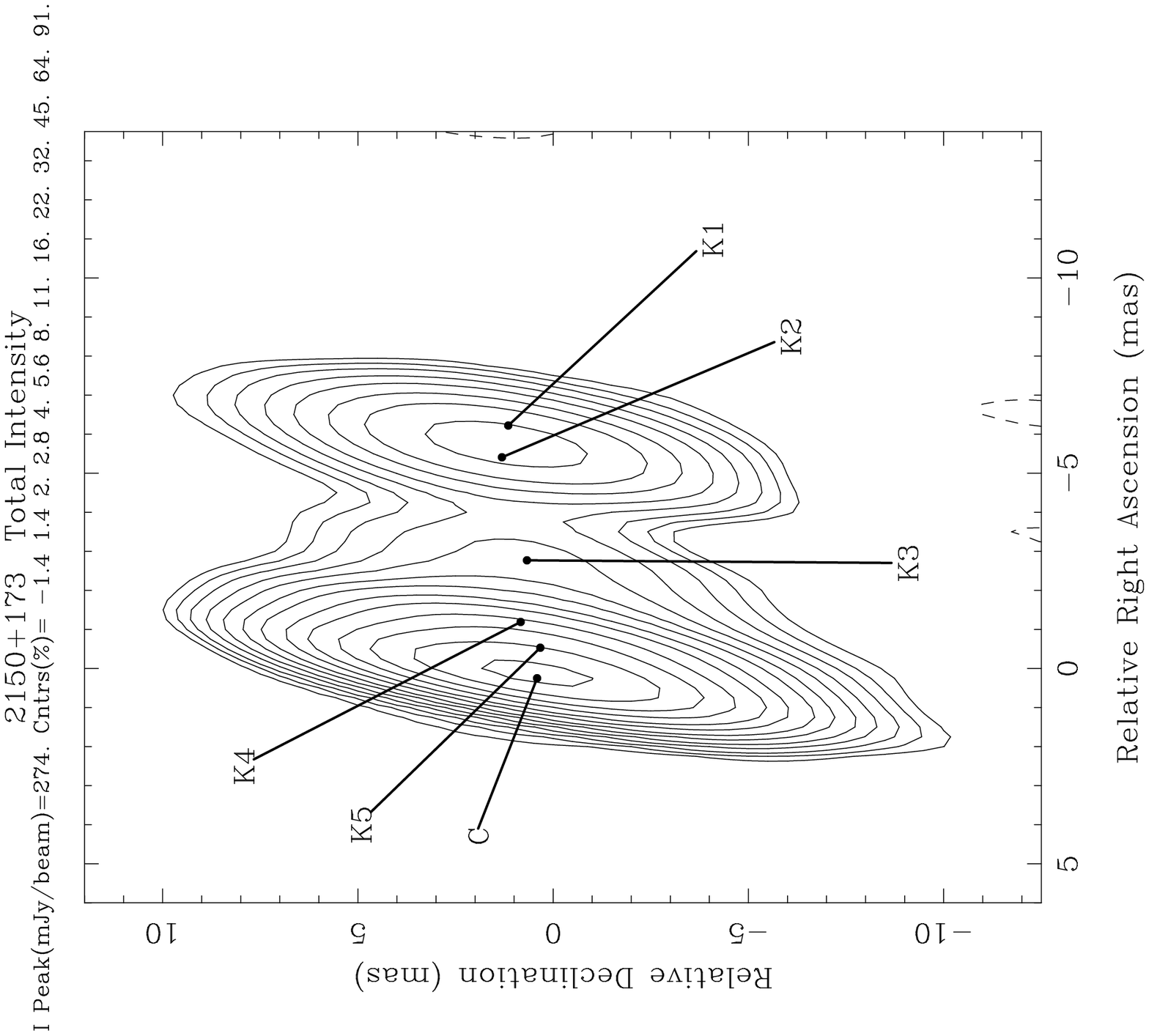,height=9.5cm}}
}
\mbox{
\rotate[r]{\psfig{file=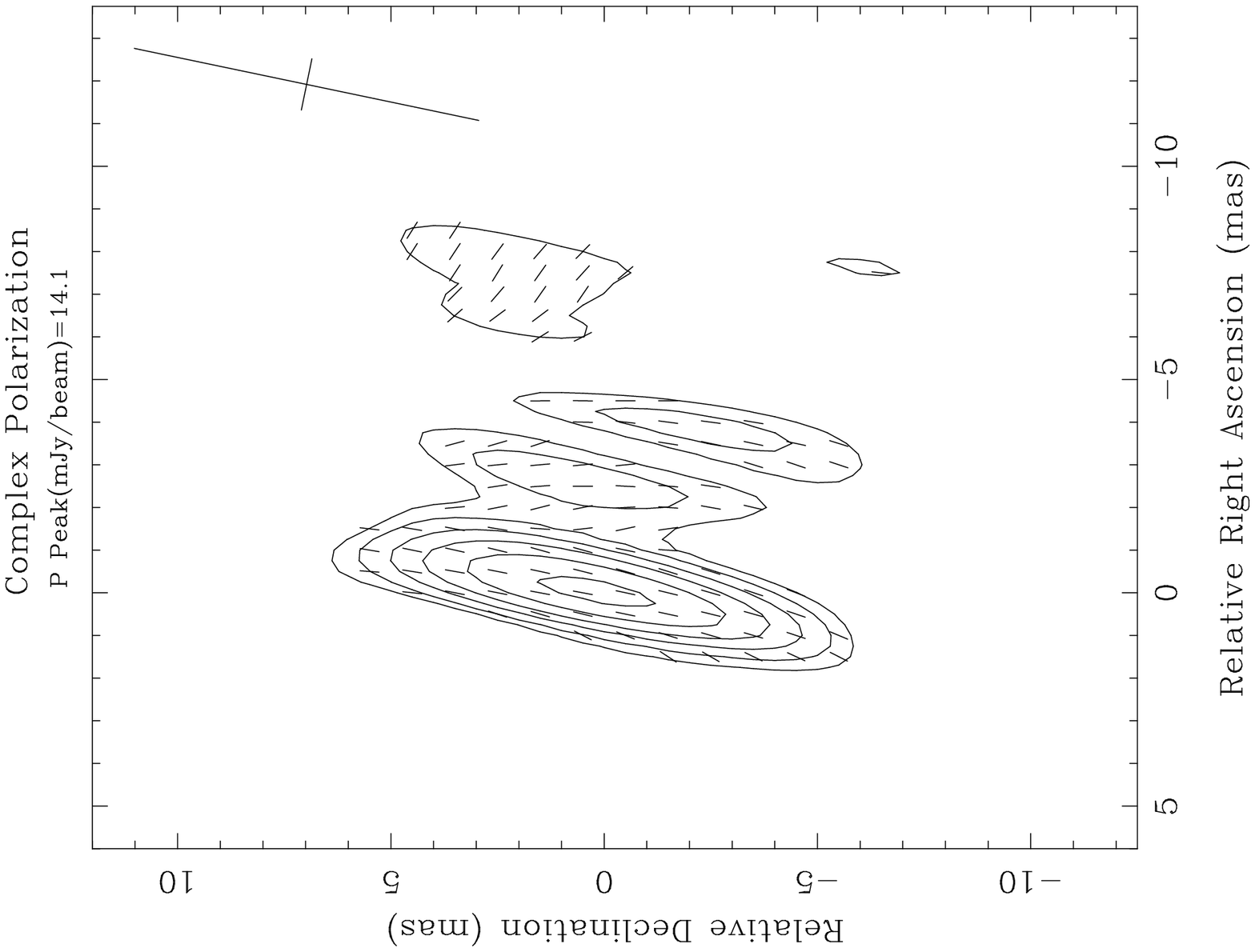,height=9.5cm}}
}
\mbox{
\rotate[r]{\psfig{file=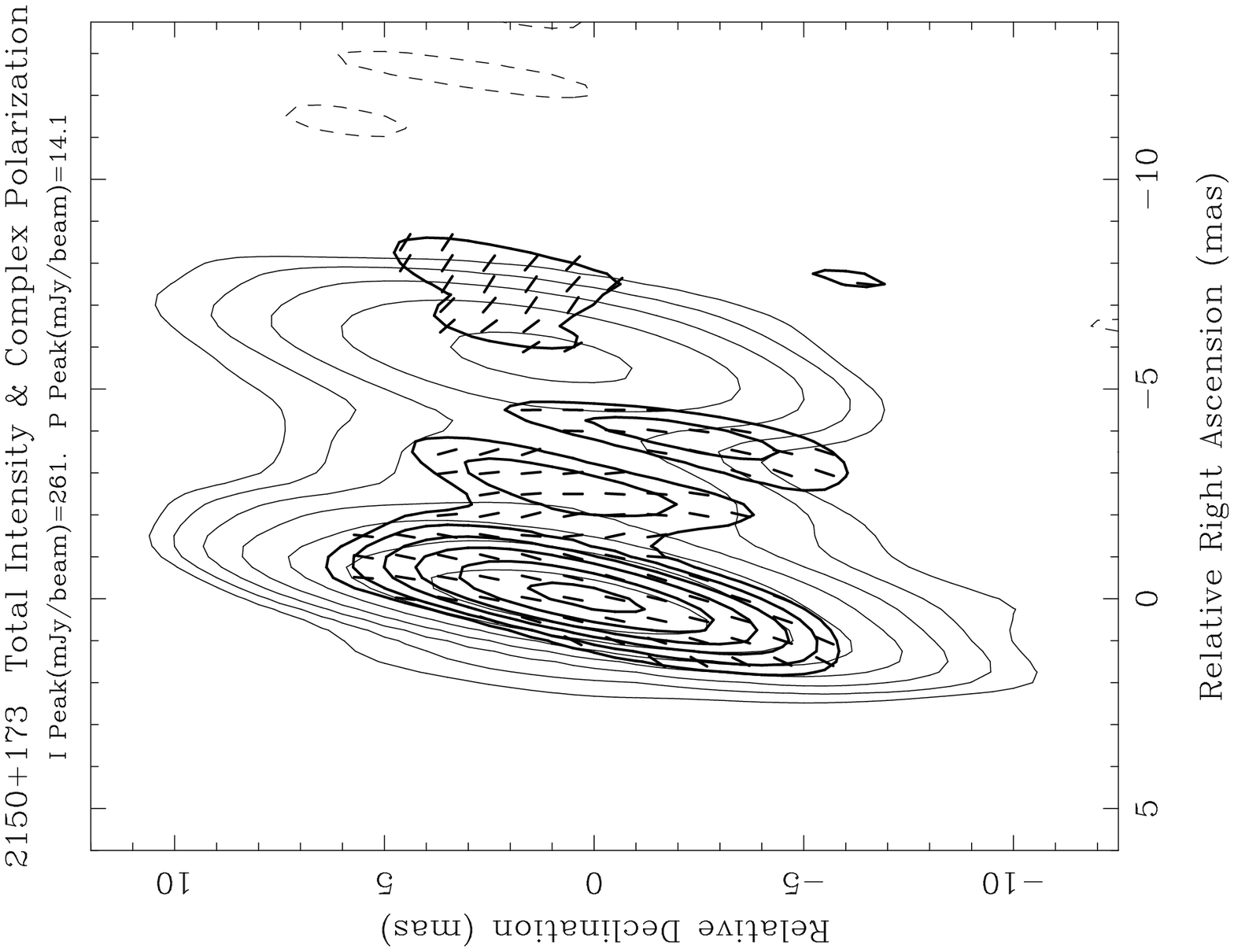,height=9.5cm}}
}
\caption{Total intensity VLBI hybrid map of 2150+173, with contours 
at $-1.4$, 1.4, 2.0, 2.8, 4.0, 5.6, 8.0, 11, 16, 23,
32, 45, 64, and 90\% of the peak brightness of 0.27 Jy/beam. (b) Linear
polarization, with contours of polarized intensity at 16, 23, 32, 
45, 64, and 90\% of the peak brightness of 14 mJy/beam, and $\chi$
vectors superimposed. (c) Superposition of $P$ image (heavy lines) over
the $I$ image, with every other $I$ contour omitted.} 
\label{fig:2150}
\end{figure}
 
\subsection{0823+033}

The redshift of this source is $z = 0.506$ (Stickel \etal 1993a). The 
2~cm image of Kellermann \etal (1998) and 13 and 4~cm images of 
Fey \& Charlot (1997) show a jet extending to the northeast. The relatively
weak jet is also visible in our images (Fig.~\ref{fig:0823}); we detected
polarization only from the VLBI core. 

\begin{figure}
\mbox{
\rotate[r]{\psfig{file=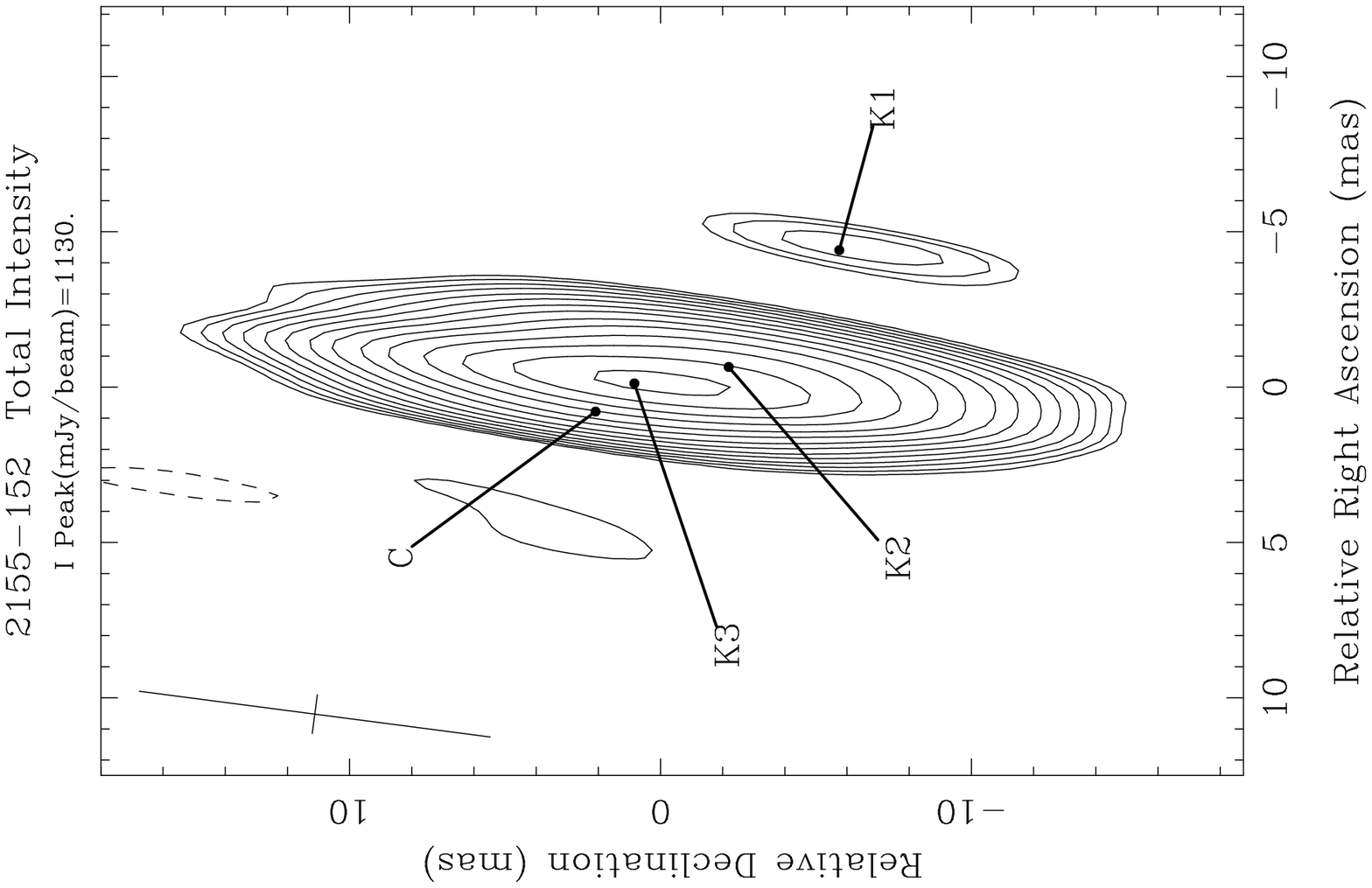,height=9.5cm}}
}
\mbox{
\rotate[r]{\psfig{file=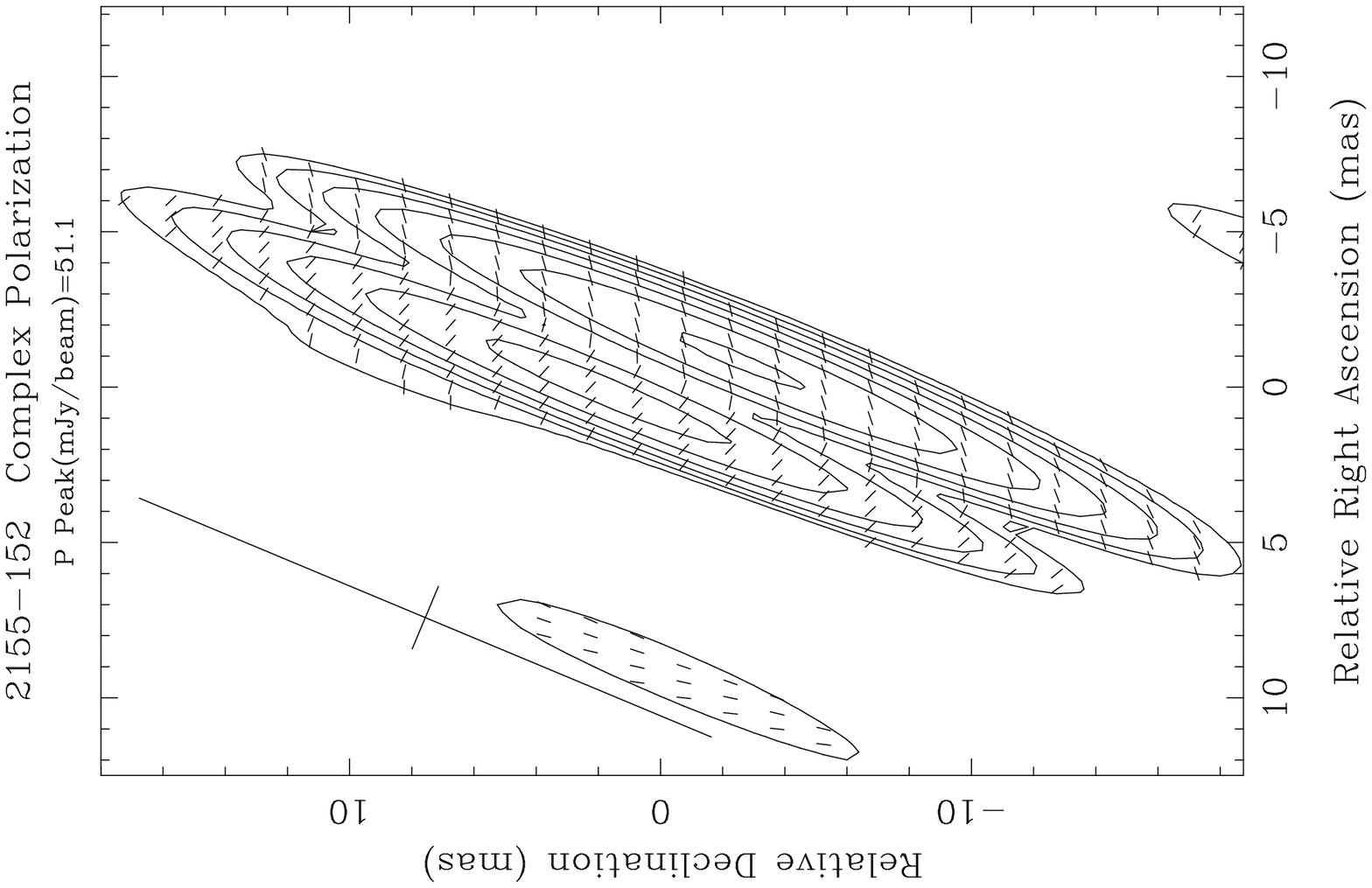,height=9.5cm}}
}
\mbox{
\rotate[r]{\psfig{file=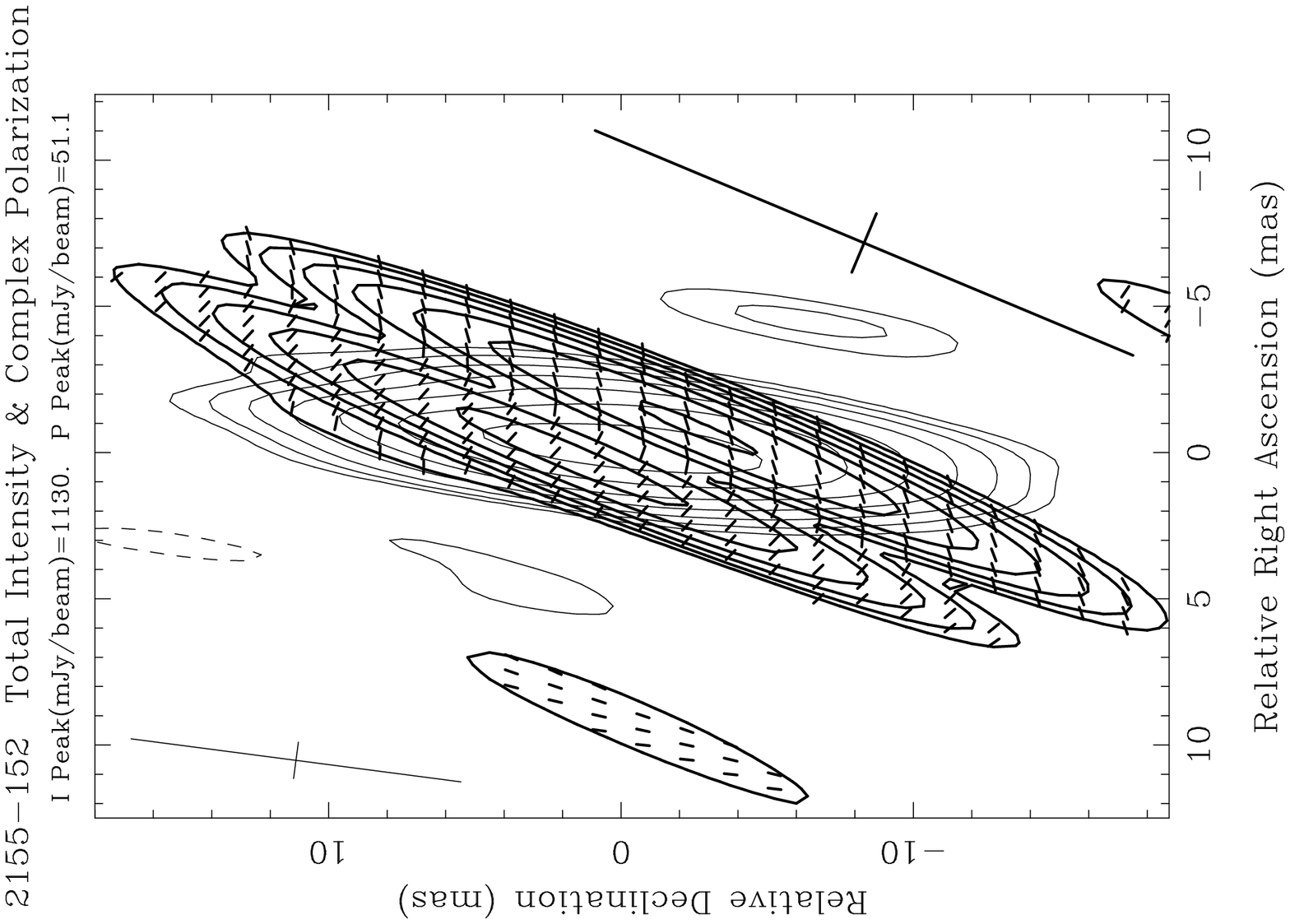,height=9.5cm}}
}
\caption{Total intensity VLBI hybrid map of 2155--152, with contours 
at $-1.0$, 1.0, 1.4, 2.0, 2.8, 4.0, 5.6, 8.0, 11, 16, 23,
32, 45, 64, and 90\% of the peak brightness of 1.22 Jy/beam. (b) Linear
polarization, with contours of polarized intensity at 17, 
24, 34, 48, 68, 96\% of the peak brightness of 51 mJy/beam, and $\chi$
vectors superimposed. (c) Superposition of $P$ image (heavy lines) over
the $I$ image, with every other $I$ contour omitted.} 
\label{fig:2155}
\end{figure}
 
\subsection{1334--127}

This extremely bright and compact source has a redshift of $z = 0.54$
(Wilkes 1986; Stickel \etal 1993b). Though this source was included
in the K\"uhr \& Schmidt (1990) BL~Lac sample, it was subsequently
reclassified as a quasar by Stickel \etal (1993b).  
The 2~cm image of Kellermann \etal (1998) shows a jet toward the
southeast. 
The dominant orientation for the inner VLBI jet in our image is 
$\theta\sim 150^{\circ}$.
The polarization of this source varied during the VLBI observations.
The $P$ map in Fig.~\ref{fig:1334}b was obtained for the time subinterval
considered in the intraday variability analysis (Gabuzda \etal 2000)
that had the best $u-v$ coverage. The polarization position angle $\chi$
in the brightest jet component is roughly perpendicular to the inner-jet 
direction. The orientation of $\chi$ in the core relative to the jet
direction is not clear; it is interesting, however, that this orientation
is nearly the same as that observed by Lister \etal (1998) in the inner 
VLBI jet at 7~mm (where it is also oblique), suggestive that our 6~cm 
``core'' emission may be associated with this same region. The 6~cm jet 
is fairly highly polarized, $\sim 18\%$. 
\subsection{1732+389}
 
The redshift of this object is $z = 0.976$, making it one of the most
distant sources in the sample. The 6~cm VLBI image of Xu \etal (1995)
showed a compact VLBI jet toward the east. Our images (Fig.~\ref{fig:1732})
show the jet more clearly. The jet contains three components,  with
polarization detected in the core and innermost jet component K3. The
$\chi$ in K3 is roughly perpendicular to the jet direction.  

\begin{figure}
\mbox{
\rotate[r]{\psfig{file=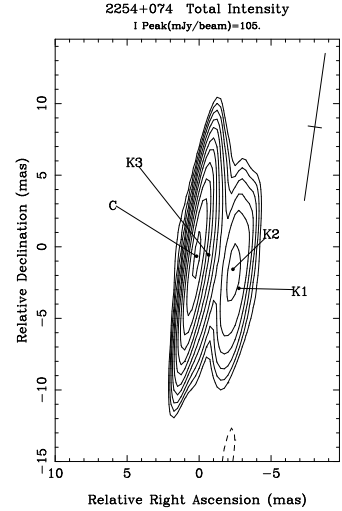,height=9.5cm}}
}
\caption{Total intensity VLBI hybrid map of 2254+074, with contours 
at $-2.8$, 2.8, 4.0, 5.6, 8.0, 11, 16, 23,
32, 45, 64, and 90\% of the peak brightness of 0.10 Jy/beam.} 
\label{fig:2254}
\end{figure}
 
\subsection{2131--021}

This source has the largest redshift among the sample 
objects, $z = 1.28$ (Drinkwater \etal 1997).  The structure is very compact,
with a jet extending nearly directly east (Fig.~\ref{fig:2131}); the
faint jet is also visible in the 2~cm image of Kellermann \etal (1998)
and the 13 and 4~cm images of Fey \& Charlot (1997). We detected
polarization only in the VLBI core, and this polarization
varied during the VLBI observations.
The $P$ map in Fig.~\ref{fig:2131}b was obtained for one time subinterval
considered in the intraday variability analysis (Gabuzda \etal 2000).
The degree of polarization was only mildly variable, $m\sim 1.6\%$,
while $\chi_{core}$ varied by about $15^{\circ}$ over the course of the
VLBI run, remaining roughly perpendicular to the direction of the VLBI
jet (to within $\sim 30^{\circ}$). 

\subsection{2150+173}

This object is one of comparatively few in the sample whose spectra are
so featureless that no redshift has been determined (Stickel \etal 1993b;
Veron-Cetty \& Veron 1993). The 13 and 4~cm VLBI images of Fey \& Charlot
(1997) show a prominent jet extending toward the northwest, which is also 
clearly visible in our images (Fig.~\ref{fig:2150}). We detected 
polarization in the core and several jet components; $\chi$ is predominantly
transverse to the jet direction in both the core and jet. 

\subsection{2155--152}

The redshift of this object is $z = 0.67$ (Stickel \etal 1989).
Its total intensity structure
(Fig.~\ref{fig:2155}) is somewhat unusual, with structure present
on both sides of the brightest feature at the phase center. We believe
that the strongest feature is a bright knot in the inner jet,
and that the feature to the northeast of this component is the core; 
multi-frequency VLBA observations currently being analysed (Pushkarev
et~al., in prep.) confirm that
the source has a core--jet structure with the jet extending toward the 
southwest. 
We detected polarization in the core to the
northeast and the second jet component to the southwest. 
The polarized flux of the core varied during the VLBI run, while
that of the jet component was constant (Gabuzda \etal 2000), with no
detectable variability in $\chi$ for either of these
components. The map in Fig.~\ref{fig:2155}b corresponds to one time 
subinterval considered in the variability analysis. $\chi$ in the core
is well aligned with the inner-jet direction, and is perpendicular
to the jet direction in the SW component, suggesting that the dominant
magnetic field in this part of the jet is longitudinal. 

\subsection{2254+074}

The host galaxy of this source, which has a redshift $z = 0.19$ 
(Stickel \etal 1988), is clearly resolved in direct images
(Stickel \etal 1993a; Falomo 1996). The BL Lac object is surrounded by a 
number of galaxies, one of which appears to be a physical companion 
(Stickel \etal 1993a). To our knowledge, no previous VLBI images are
available. We detected three components
in the VLBI jet (Fig.~\ref{fig:2254}); the polarization of this source was 
not sufficiently strong to be detected. 

\section{VLBI properties of a complete sample of 1-Jy BL Lacertae 
objects at 6~cm}

\begin{figure}
\mbox{
\rotate[r]{\psfig{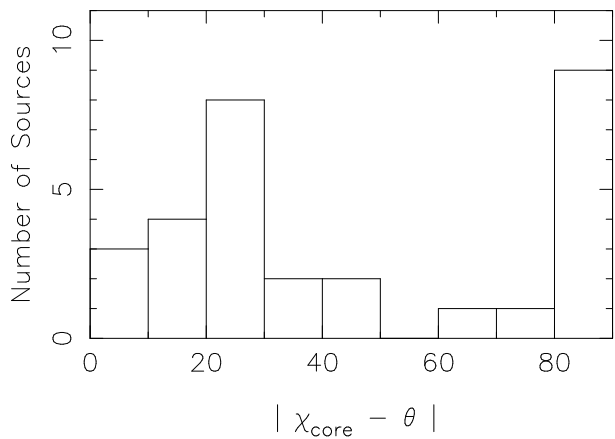}}
}
\caption{Distribution of offsets between the core $\chi$ value (after
correction for the integrated rotation measure) and the direction of
the inner VLBI jet $|\chi_{core}-\theta|$ for the 23 
K\"uhr \& Schmidt (1990) BL Lac objects for which such measurements
are available. The distribution is clearly bimodal, with preferred
values close to $0^{\circ}$ and $90^{\circ}$.}
\label{fig:corehist}
\end{figure}
 
\begin{figure}
\mbox{
\rotate[r]{\psfig{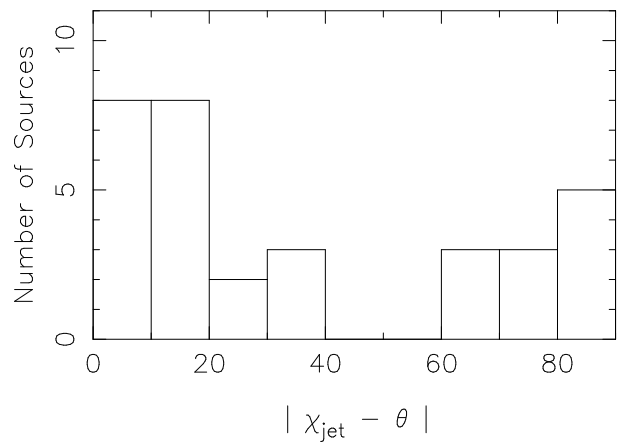}}
}
\caption{Distribution of offsets between the jet $\chi$ value (after
correction for the integrated rotation measure) and the local VLBI
jet direction $|\chi_{jet}-\theta|$ for the 20 
K\"uhr \& Schmidt (1990) BL Lac objects for which such measurements
are available. The distribution shows a clear predominance of values 
close to $0^{\circ}$, with a weaker secondary peak near $90^{\circ}$.}
\label{fig:jethist}
\end{figure}
 
Having completed our first-epoch 6~cm polarization observations of
all sources in the K\"uhr \& Schmidt sample of BL Lacertae objects, 
we are now able to evaluate the VLBI properties of the sample as
a whole. We summarize our results for the VLBI polarization properties
of the sample sources in Table~4. The columns present
(1) Source name (1950.0), (2) redshift $z$, (3) Epoch for the VLBI
polarization observations, (4) total flux density on VLBI scales $I_{VLBI}$,
(5) fraction of VLA
core flux present on VLBI scales $f_I$, (6) total polarized flux
density on VLBI scales $p_{VLBI}$, (7) fraction of VLA core polarized 
flux present on VLBI scales $f_P$, (8) position angle for the
total polarization on VLBI scales $\chi_{VLBI}$, (9) position angle for 
the polarization of the VLA core $\chi_{VLA}$,
(10) degree of polarization in the VLBI core $m_c$, (11) degree of
polarization in the VLBI jet $m_j$, (12) offset between $\chi_{core}$
and the jet direction $|\theta-\chi_{core}|$, (13) offset between
$\chi_{jet}$ and the jet direction $|\theta-\chi_{jet}|$, (14)
integrated rotation measure RM, and (15) references. Separate rows are
given for each epoch for which VLBI polarization data have been analyzed. 
When polarization was detected in more than one jet component at a given
epoch, multiple entries separated by commas are given. 

The $\chi_{VLBI}$
and $\chi_{VLA}$ values are as observed, i.e., not corrected for
Faraday rotation, but $|\theta-\chi_{core}|$ and $|\theta-\chi_{jet}|$
have both been determined after correction for the integrated rotation
measure listed in column (16). The one exception is 0820+225, for
which multi-frequency VLBI polarization observations have revealed
a non-uniform distribution of the rotation measure on milliarcsecond
scales (Gabuzda, Pushkarev \& Garnich, in prep.); in this case, the values 
in columns (14) and (15) have been determined after correction for the 
local rotation measure in the VLBI jet. 

\subsection{Properties of the VLBI core polarization}

Previous observations (Gabuzda \etal 1994b, 1999, and references
therein) have indicated that the VLBI core polarizations of BL Lac
objects are typically $\sim 2-5\%$. After completing our first-epoch
polarization observations for the entire K\"uhr \& Schmidt BL Lac
sample, we have measurements of the VLBI core polarizations $m_c$ for all
but five of the 33 sample sources. These $m_c$ values, collected in
Table~4, with few exceptions range from 1.8 to 6.8\%, 
demonstrating that the behaviour suggested by the earlier incomplete
observations is, indeed, characteristic of BL~Lac objects. 

Another tendency indicated by the earlier observations was for the
core polarization position angle $\chi_{core}$ to be either aligned
with or transverse to the direction of the inner 6~cm VLBI jet. We
now have measurements of $|\theta - \chi_{core}|$ for 24 of the 33
sample sources. The
$|\theta - \chi_{core}|$ values in Table~4 and in
Fig.~\ref{fig:corehist} clearly show this behaviour: in 16 cases, 
$|\theta - \chi_{core}|$ is between 0 and $30^{\circ}$, and 
in 12 cases $|\theta - \chi_{core}|$ is between 60 and $90^{\circ}$. This
indicates that the $|\theta - \chi_{core}|$ distribution for this
$\sim 73\%$ of the sample is bimodal, with the numbers of sources
displaying either of the two preferred values for $|\theta - \chi_{core}|$
being roughly equal. It appears that the peak at small misalignments
may be offset from $0^{\circ}$; the origin of this behaviour is not
clear, though it could be associated with uncertainty in the jet
direction on scales slightly smaller than the resolution of our 6~cm
global observations. 

\subsection{Characteristics of the jet magnetic fields}
 
Earlier observations indicated a tendency for $\chi$ in the VLBI jets of 
BL Lacertae objects to be parallel to the local jet direction, which 
has usually been interpreted as evidence for the presence of relativistic 
plane shocks. We now have measurements of $|\theta - \chi_{jet}|$ for 25 
of the 33 sample sources. The $|\theta - \chi_{jet}|$ values in 
Table~4 and in Fig.~\ref{fig:jethist} clearly show a predominance of 
sources with $\chi_{jet}$ and $\theta$ aligned to within $\sim 30^{\circ}$ 
($\sim 60\%$). At the same time, the histogram shows a weaker secondary 
peak of sources with $|\theta - \chi_{jet}|$ near $90^{\circ}$. 
Thus, it is clear that, as hinted by previous observations (Gabuzda \etal
1999 and references therein), a sizeable fraction ($\sim 30\%$) of sources 
in the K\"uhr \& Schmidt (1990) sample of BL Lacertae objects have 
longitudinal magnetic fields in their VLBI jets. At this time, it is
not clear how jet components with transverse and longitudinal magnetic
fields physically differ; multi-frequency VLBA polarization observations
for all the sample sources currently under analysis will hopefully 
help eludicate this question in the near future.

In the remaining sources, $\chi_{jet}$ bears no obvious relation to 
$\theta$, possibly due to the presence of Faraday rotation with 
substantially different values than suggested by the integrated rotation 
measures. Multi-frequency polarization observations currently being
analyzed should shed light on this issue. 

\subsection{Superluminal motion}

Among the sources for which images are presented and analysed here, 
to our knowledge, previous VLBI images at 6~cm or nearby wavelengths
are available for five of them:
0003--066, 0823+033, 1732+389, 2131--021, and 2150+173. 1732+389
was observed at epoch 1990.18 as part of the 6~cm Caltech--Jodrell Bank
Survey (Xu \etal 1995). The other four sources were observed at 4~cm by
Fey \& Charlot (1997; hereafter FC). 

In their 6-cm image of 1732+389, Xu \etal (1995) detected a single
jet component at $r = 0.54$~mas. It is natural to tentatively identify
this feature with K3 in Table~\ref{tab:models}. Our second-epoch
observations at epoch 1995.41 support this identification, and indicate
that the apparent proper motion of this feature is 0.24 mas/yr, which
corresponds to an apparent speed $\beta_{app}h = 6.5$ (Pushkarev \&
Gabuzda 1999a). The feature K2 is also present in our 1995.41 image;
a comparison suggests a tentative proper motion of 0.61~mas/yr,
corresponding to an apparent speed $\beta_{app}h = 16.5$ (Pushkarev \&
Gabuzda 1999a).

FC obtained images of a large number of compact 
extragalactic radio sources at 13 and 3.6~cm. We will be interested
primarily in the 3.6~cm images, since they have higher resolution and
the observing wavelength is closer to that for our data. We must be 
aware of possible small frequency-dependent
shifts in the apparent separations of jet components from the core when 
comparing these 3.6~cm images with our own 6~cm images. However, these two
wavelengths are close enough that we would not expect such shifts to
be significant in most cases, so that the 3.6~cm data of FC
can be used to derive at least tentative estimates of the structural
changes in the sources observed in both projects. Below, we will refer 
to the 3.6~cm images of FC unless explicitly stated
otherwise. 

FC's 3.6~cm observations of 0003--066 were less than five months after our
first-epoch observations. There is a reasonably close correspondance
between the jet structures detected in the two observations, but the
interval between them is too short to reliably identify systematic proper 
motions. Most features appear to be stationary to within the errors.
We therefore postpone estimation of the apparent speeds in
the VLBI jet of this source until the results of our second-epoch
observations become available.  

\begin{figure}
\mbox{
\rotate[r]{\psfig{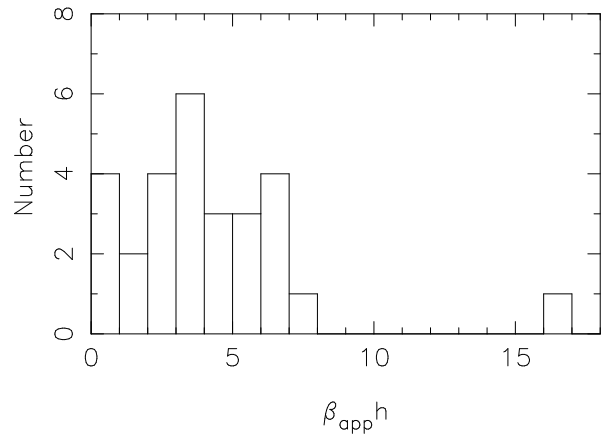}}
}
\caption{Distribution of apparent superluminal speeds for the sixteen
K\"uhr \& Schmidt (1990) BL Lac objects with redshifts for which
component speed estimates are available. With the exception of the
one very high speed of $\beta_{app}h = 16.6$, the range of apparent 
speeds is relatively small, and extends to considerably lower speeds
than observed in core-dominated quasars (Fig.~\ref{fig:speeds_cdq}).}
\label{fig:speeds}
\end{figure}
 
\begin{figure}
\mbox{
\rotate[r]{\psfig{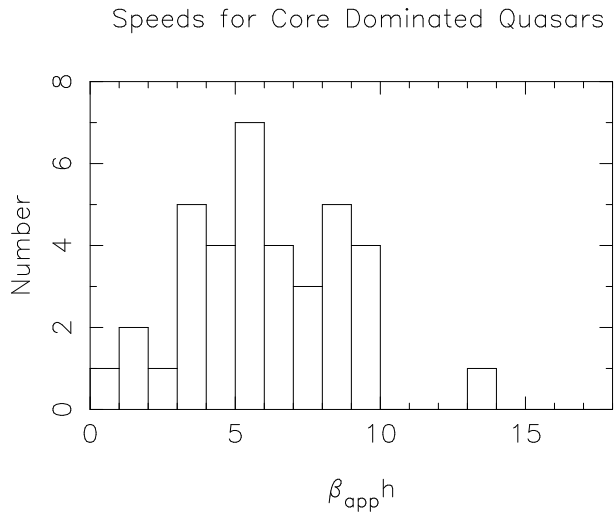}}
}
\caption{Distribution of apparent superluminal speeds for  
nineteen core-dominated quasars from Vermeulen \& Cohen (1994). 
The distribution extends to appreciably higher speeds than the
distribution for the K\"uhr \& Schmidt BL Lacertae objects shown in
Fig.~\ref{fig:speeds}.}
\label{fig:speeds_cdq}
\end{figure}
 
FC detected only a single jet component in 0823+033
at epoch 1995.28, at $r = 1.9$~mas. There are no components in our image 
near this position. If we assume expansion of the source structure with
time, the simplest joint interpretation of the two images is that the
feature detected by FC is K3 in Table~\ref{tab:models}.
In this case, its tentative proper motion is 0.39~mas/yr, which
corresponds to an apparent speed $\beta_{app}h = 6.9$. 

Two jet components were detected by FC in 2131--021: one at $r = 0.5$~mas
in $\theta = 107^{\circ}$ and another at $r = 1.6$~mas in $\theta = 
75^{\circ}$. A comparison with the model for this source in 
Table~\ref{tab:models} shows that these positions nearly coincide with
those for our components K5 and K4. In both cases, the positions of
these features in the two images agree to within the errors. Thus, the
simplest joint interpretation of the two images is that K4 and K5 were
stationary from epoch 1992.23 to epoch 1995.78.

There is a very good correspondance between the structures in our
image of 2150+173 and the image of FC, in which features were detected
at $r = 0.8, 1.9, 4.0,$ and 6.9~mas. It seems likely that these four
components should be identified with K5, K4, K3, and K2 in 
Table~\ref{tab:models}. In this case, the implied proper motions for
these features are $\mu = 0.01, 0.11, 0.27,$ and 0.32~mas/yr, respectively; 
we can see that this suggests acceleration with distance from the core,
but further observations are required to test this hypothesis. Unfortunately,
the redshift of this source is unknown, so that it is not possible to
directly translate these proper motions into apparent speeds. 

Finally, we have one more tentative speed to report here. 
At the time of the analysis for our first-epoch 6~cm observations of 0138--097
(Gabuzda \etal 1999), we were not aware that FC had
also observed this source at 3.6~cm. They detected a single jet component
1.9~mas from the core. If this feature should be identified with K2 from
Gabuzda \etal (1999), this implies a proper motion of 0.35 mas/yr, which
corresponds to a tentative apparent speed $\beta_{app}h = 5.53$.

Our observations have substantially increased the number of objects in 
the K\"uhr \& Schmidt sample with at least tentative (two-epoch) speed 
estimates. Such estimates are now available for 16 of the 33 sample sources,
and are summarized in Table~\ref{tab:speeds}. The 
histogram in Figure~\ref{fig:speeds} shows the distribution of superluminal 
speeds in these 16 objects. When making
this histogram, we included more than one entry per source only if apprecially
different speeds were observed for different components (i.e., only if
the speeds fell in different bins). We can see that
there is a concentration toward relatively low values, and the apparent 
speeds are fairly uniformly distributed in the range from $0-7\beta_{app}h$,
with a possible weak peak near $\sim 4$.  We see no evidence from this 
distribution that the typical jet Lorent factors exceed $\gamma \sim 5-6$. 
There is some evidence for the occurrence of higher superluminal speeds 
in some BL Lacertae objects on the smaller scales probed by higher-frequency 
observations (e.g., Marscher \& Marchenko 1998); it remains unclear whether 
the jets of BL Lacertae objects in general tend to display higher superluminal 
speeds on smaller scales. 

Figure~\ref{fig:speeds_cdq} shows the distribution of superluminal speeds
for components in 19 core dominated quasars taken from Vermeulen \&
Cohen (1994). Here, as for the distribution in Fig.~\ref{fig:speeds},
we have not plotted more than one entry per source per bin. We can see 
that the quasar speed distribution has fewer
values below $\sim 3$ and appreciably more values exceeding $\sim 7$ than
the BL Lac distribution.  The overall centroid of the distribution is 
shifted toward higher speeds: the median and average speeds for the
BL Lacertae objects are both 3.7, while the corresponding values for the
quasars are both 5.8.
A Kolmogorov--Smirnov test indicates that the probability that the two
distributions are the same is less than 1\%. This confirms previous 
evidence that the superluminal speeds in BL Lac objects are systematically 
lower than those observed in core-dominated quasars (Gabuzda \etal 1994b; 
Britzen \etal 1999). 

\begin{table}
\addtocounter{table}{1}
\caption[xx]{{\sc Component Speeds in K\"uhr \& Schmidt BL Lacertae Objects}}
\label{tab:speeds}
\begin{center}
\begin{tabular}{cccccc} \hline \hline
Source    & $z$      & N & $\mu$ & $\beta_{app}h$ & Ref. \\
          &          &   & (mas/yr) &             &       \\\hline
0138--097 & 0.44     & 2 & 0.35     & 5.5         & 12 \\
0454+844  & 0.11     & 5 & 0.14     & 0.6         & 1,3\\
0716+714  & $\ldots$ & 4 & 0.07     & $\ldots$    & 1,8\\
          & $\ldots$ & 2 & 0.11     & $\ldots$    & 8\\
0735+178  & $>0.42$  & 4 & 0.48     & 7.4         & 5.6\\
          &          & 4 & 0.33     & 5.0         & 6\\
          &          & 3 & 0.28     & 4.2         & 6\\
0823+033  & 0.51     & 2 & 0.39     & 6.9         & 12 \\
0828+493  & 0.55     & 2 & 0.34     & 6.3         & 9 \\
0851+202  & 0.31     & 3 & 0.20     & 2.4         & 2\\
          &          & 3 & 0.27     & 3.2         & 2\\
0954+658  & 0.37     & 2 & 0.37     & 5.2         & 7\\
          &          & 2 & 0.44     & 6.2         & 7\\
1418+546  & 0.15     & 3 & 0.56     & 3.6         & 9,10\\
          &          & 2 & 0.08     & 0.5         & 9,10\\
1652+398  & 0.03     & 3 & 0.27     & 0.4         & 7\\
          &          & 2 & 0.55     & 0.8         & 7\\
1732+389  & 0.97     & 3 & 0.24     & 6.5         & 10,12 \\
          &          & 2 & 0.61     & 16.6        & 10,12 \\
1749+701  & 0.77     & 3 & 0.14     & 3.2         & 7\\
          &          & 3 & 0.08     & 2.0         & 7\\
          &          & 2 & 0.09     & 2.1         & 7\\
1803+784  & 0.68     & 2 & 0.08     & 1.8         & 7\\
1823+568  & 0.34     & 3 & 0.18     & 3.8         & 7\\
          &          & 2 & 0.20     & 4.3         & 7\\
2007+777  & 0.34     & 5 & 0.22     & 2.9         & 1,7\\
2200+420  & 0.07     & 4 & 1.14     & 3.7         & 4\\
          &          & 5 & 1.12     & 3.6         & 4\\
          &          & 4 & 1.12     & 3.6         & 4\\
          &          & 2 & 1.00     & 3.3         & 4\\ 
2150+173  & $\ldots$ & 2 & 0.01     & $\ldots$    & 12 \\ 
          &          & 2 & 0.11     & $\ldots$    & 12 \\ 
          &          & 2 & 0.28     & $\ldots$    & 12 \\ 
          &          & 2 & 0.34     & $\ldots$    & 12 \\ 
2254+074  & 0.19     & 2 & 0.46     & 3.7         & 11,12 \\ 
          &          & 2 & 0.30     & 2.4         & 11,12\\ 
          &          & 2 & 0.55     & 4.3         & 11,12 \\ \hline\hline
\multicolumn{6}{l}{Refs: 1 = Witzel \etal 1988; 2 = Gabuzda, Wardle \& Roberts
1989a;}\\
\multicolumn{6}{l} {3 = Gabuzda \etal 1989; 4 = Mutel \etal 1990; 5 = }\\
\multicolumn{6}{l} {B{\aa\aa}th \& Zhang 1991; 6 = Gabuzda \etal 1994a;}\\
\multicolumn{6}{l} {7 = Gabuzda \etal 1994b; 8 = Gabuzda \etal 1998;}\\
\multicolumn{6}{l} {9 = Gabuzda \etal 1999; 10 = Pushkarev \& Gabuzda 1999a;}\\
\multicolumn{6}{l} {11 = Pushkarev \& Gabuzda 1999b; 12 = This paper.}\\
\end{tabular}
\end{center}
\end{table}

\section{Conclusion}

This paper completes analysis of our first-epoch observations of all 
sources in the complete sample of BL Lacertae objects defined by K\"uhr \&
Schmidt (1990). Certain tendencies were present even in the earliest
VLBI polarization observations of BL Lac objects; however, without systematic
observations of a complete sample of objects, it was not possible to say
whether these properties were relevant for radio-loud BL Lac objects in 
general, or only to a handful of the best studied sources. On the whole, 
our analysis of the complete sample has confirmed the results of these 
early observations. However, it is also clear that the behaviour shown
by the sample sources is not entirely uniform. We summarize our results
below.

The VLBI core polarizations of BL Lacertae objects are appreciable, with
values typically ranging from $\sim 2-7\%$, and occasionally reaching values
as high as $\sim 10\%$. Gabuzda \etal (1994b) suggested that these relatively
high values reflected the dominant contribution of newly emerging jet 
components. 
In the case of one source in the sample -- 1803+784 -- 6~cm space VLBI 
polarization observations directly showed that the polarized flux was 
dominated by a compact component in the inner VLBI jet, which was unresolved 
from the core in ground-based observations (Gabuzda 1999). 

The 6~cm VLBI core polarizations of quasars are much lower, typically 
$\leq 2\%$ (Cawthorne \etal 1993). This suggests that either the cores
of quasars are depolarized (see, e.g., Taylor 1998, 2000), or that they 
are considerably less likely to 
be dominated by emission from compact new jet components. If the VLBI
cores of quasars are depolarized, we expect that the observed degrees of 
polarization of the cores of quasars and BL Lac objects will become more 
similar at higher frequencies; thus far, there is no clear evidence for
this, but more systematic studies are required.

Thus, it remains a possibility that the VLBI core polarizations of BL Lac 
objects are more often dominated by the contribution of emerging jet 
components than are the VLBI core polarizations of quasars. The origin of 
this systematic difference is not obvious. One possibility is that the birth 
rate for new jet components is higher in BL Lacertae objects than in 
quasars, increasing the probability of observing a BL Lac object core 
harbouring new jet components.  There is some evidence from the University 
of Michigan monitoring database that outbursts in BL Lacertae objects may 
be more frequent and well resolved than those in quasars, suggesting that 
BL Lac objects may generate new components more frequently (M. Aller, 
private communication). Since the superluminal speeds observed in 
BL Lacertae objects are, on average, slower than those in quasars, 
it may be that the jet components of BL Lacertae objects 
spend more time in the unresolved core region before becoming detectable 
as distinct jet knots. Another possibility is that highly-polarized jet 
components close to the core are often long-lived
stationary components whose polarization blends with the true core
polarization. Quasi-simultaneous multi-frequency VLBA observations for all 
sources in the K\"uhr \& Schmidt sample currently being analyzed should 
help distinguish between these various possibilities. 

One of the striking tendencies noted in the earliest VLBI
polarization results was for the jets of BL Lacertae objects to
have transverse magnetic fields. Our first-epoch images for the complete 
K\"uhr \& Schmidt sample confirm that this is the predominant behaviour 
for the sample as a whole: among the 25 sources in which jet polarization was
detected, some 60--70\% have transverse magnetic fields. At the same time,
a sizeable minority of about 30\% have longitudinal jet magnetic fields.
Thus far, transverse magnetic fields have usually been interpreted as
manifestations of relativistic shocks in the VLBI jets of these sources,
while longitudinal fields have been taken to reflect the presence of
shear between the jet and the surrounding medium. In this picture, the
common presence of transverse magnetic fields reflects the existence of
conditions favorable for the formation of transverse shocks; the jet
components with dominant longitudinal fields would be those in which
shocks did not form, or did form but were dominated by the effect of shear. 

Though it seems likely that many individual, compact, highly-polarized
features with transverse magnetic fields are associated with 
shocks, this does not necessarily imply that {\em all} the observed
transverse jet fields should be identified with shock components. Another 
possibility is that, in at least some cases, we are detecting the 
toroidal component of an intrinsic helical jet magnetic field (see, e.g.,
Gabuzda 1999). 
If the dominant magnetic field in the jet is helical, the net observed
field can be either transverse or longitudinal, depending on the pitch angle
of the field and the viewing angle, though it is more likely that the
net observed field will be transverse. In addition, a longitudinal field
component could develop due to interaction between the edges of the jet
and the surrounding medium (e.g. Aaron 1998, Laing \etal 1999, Aloy \etal 
2000).  Thus, it could be that the dominant magnetic-field component in 
the VLBI jets of BL Lacertae objects is characteristically toroidal. 
This could be consistent with
the fact that the observed jet magnetic fields are most often transverse,
but occasionally longitudinal. 

The characteristically modest superluminal speeds observed in the K\"uhr \&
Schmidt sources suggest that BL Lacertae objects differ from quasars in
either the characteristic angles of their jets to the line of sight, or
the characteristic intrinsic velocities of components in their jets, or both.
The observed apparent speed for a VLBI feature has a peak for motion at
an angle to the line of sight of about $\theta\sim 1/\gamma$, where $\gamma$
is the Lorentz factor of the motion. Therefore, if the intrinsic velocities
in the two types of sources were essentially the same, the BL Lacertae objects
could, in principle, have smaller apparent velocities if their jets were
significantly further from or nearer to the line of sight than the jets in
quasars. Since BL Lacertae objects are obviously highly beamed sources
(Kollgaard 1994, for example),
it is not reasonable to suppose that their jets could typically be at
significantly larger angles to the line of sight than those in quasars. On
the other hand, if their jets were significantly {\it closer} to the line of
sight than quasar jets, we would expect BL Lacertae objects to be significantly
more highly beamed than quasars, and there is no evidence for this
(Ghisellini \etal 1993).
Thus, the most straightforward interpretation of the more modest superluminal
speeds observed in the VLBI jets of BL Lacertae objects compared to quasars
is that the intrinsic velocities in the BL Lac jets are lower. This seems
quite natural in the context of unified schemes linking BL Lacertae
objects with FR~I and quasars with FR~II radio galaxies. 

\section{Acknowledgements}

This work was supported by an American Astronomical Society Henri Chretien 
International Research Grant and Small Grant (DCG). ABP acknowledges  
support from the Russian Foundation for Basic Research. DCG acknowledges
support from the European Commission under the IHP Programme (ARI) contract No.
HPRI-CT-1999-00045. We would like to thank R. M. Campbell for useful 
discussions of some aspects of our analysis. We also thank the staff at 
the participating observatories and correlation facillities who made these 
observations possible.  This
research has made use of data from the University of Michigan Radio Astronomy
Observatory, which is supported by the National Science Foundation and by
funds from the University of Michigan.

\end{document}